\documentclass[11pt,a4paper]{article}
\pdfoutput=1
\usepackage{jheppub}

\newcommand{\be}{\begin{equation}}
\newcommand{\ee}{\end{equation}}
\newcommand{\bea}{\begin{eqnarray}}
\newcommand{\eea}{\end{eqnarray}}

\numberwithin{equation}{section}
\usepackage{bm}

\title{\boldmath Coexistence of two vector order parameters: a 
holographic model for ferromagnetic superconductivity}

\author[a]{Andrea Amoretti,}
\author[b]{Alessandro Braggio,}
\author[a]{Nicola Maggiore,}
\author[a]{Nicodemo Magnoli}
\author[c]{and Daniele Musso}

\affiliation[a]{Dipartimento di Fisica, Universit\`a di Genova,\\
via Dodecaneso 33, I-16146, Genova, Italy\\and\\I.N.F.N. - Sezione di Genova\\}
\affiliation[b]{CNR-SPIN, Via Dodecaneso 33, 16146, Genova, Italy\\}
\affiliation[c]{Physique Th\'eorique et Math\'ematique\\
Universit\'e Libre de Bruxelles, C.P. 231, 1050 Bruxelles, Belgium}

\emailAdd{andrea.amoretti@ge.infn.it}
\emailAdd{alessandro.braggio@spin.cnr.it}
\emailAdd{nicola.maggiore@ge.infn.it}
\emailAdd{nicodemo.magnoli@ge.infn.it}
\emailAdd{dmusso@ulb.ac.be}

\abstract{We study a generalization of the standard holographic p-wave 
superconductor featuring two interacting vector order parameters.
Basing our argument on the symmetry and linear response properties of the
model, we propose it as a holographic effective theory describing
a strongly coupled ferromagnetic superconductor. We show that 
the two order parameters undergo concomitant condensations as a manifestation of
an intrinsically interlaced charge/spin dynamics. Such intertwined 
dynamics is confirmed by the study of the transport properties. We characterize 
thoroughly the equilibrium and the linear response (i.e. optical conductivity and spin susceptibility)
of the model at hand by means of a probe approximation analysis. Some insight
about the effects of backreaction in the normal phase can be gained by analogy with the s-wave
unbalanced holographic superconductor.}

\begin{document}

\maketitle

\section{Introduction}

Gauge/gravity correspondence has provided us with a novel theoretical framework to investigate 
the strongly coupled regime of quantum field theory \cite{Aharony:1999ti}. 
Such fruitful approach has a particularly natural application to quantum critical and strongly 
correlated systems which are ubiquitous in condensed matter \cite{Sachdev:2010ch,Hartnoll:2009sz,Herzog:2009xv}. 
In this context, the models featuring spontaneous symmetry breaking allowed 
new investigations about the physics of high $T_c$ superconductors (HTC)  \cite{Gubser:2008px,Hartnoll:2008vx,Hartnoll:2008kx}.
The present model belongs to such a family.
Indeed, we propose a bottom-up holographic model to account for the interplay of two strongly coupled 
vector order parameters. The phenomenological purpose of describing a ferromagnetic superconductor
led us to generalize the standard holographic p-wave model introduced in \cite{Gubser:2008wv}.
The generalization consists in adding a second non-Abelian gauge field in the bulk and considering 
direct interactions between the two vector fields. Such direct interactions are encoded in a term that
couples the field strengths of the two gauge fields constraining them to transform according to the 
same gauge transformation. The diagonalization of the 
bulk kinetic terms leads to an interesting field structure featuring the interaction between 
a gauge field and massless vectorial matter in the adjoint representation of the gauge group. 

The present model belongs both to the class of holographic systems containing spinful order parameters 
and to the class possessing more than one order parameter. In the former group, besides the p-wave superconductors,
we can also find models describing d-wave order parameters \cite{Benini:2010pr}. In the latter class
we find either models describing multi-band superconductors \cite{Huang:2011ac,Krikun:2012yj} or models 
which investigate the coexistence of different orderings \cite{Basu:2010fa,Musso:2013ija,Cai:2013wma,Nie:2013sda,Amado:2013lia}%
\footnote{Models with more than one complex scalar order parameter have also been studied in the context
of holographic analysis of type II Goldstone bosons \cite{Amado:2013aea,Amado:2013xya}.}.

\subsection{Motivations}

The main aim motivating the model studied here is to describe 
the coexistence and interactions of two vector order parameters at strong coupling
in a holographic framework.
Investigations about the interplay of multiple orderings are relevant in many areas 
of physics ranging from particle physics to condensed matter. For example, spin density waves in cuprates 
(or, more in general, magnetic orderings and superconductivity), multi-condensate superconductors but also magnetism in 
neutron stars. In relation to particle physics, an example is provided by models having 
a Higgs-doublet%
\footnote{To have a wider list and an account of how the subject is treated within 
the standard Ginzburg-Landau effective theory paradigm, see \cite{Ivanov:2009} and 
references therein.}.

Albeit the phenomenological interpretation on which we focus
is mainly concerned with the condensed matter panorama, it is important to pinpoint that the 
present holographic model could have in principle a wider applicability. 
The subject regarding the coexistence of different orderings has been already addressed in the holographic literature 
and one of our main purposes here is to extend the study 
presented in \cite{Musso:2013ija} where a minimal holographic model with two scalar orderings 
was thoroughly analyzed.
More precisely, we consider a generalization of the p-wave holographic superconductor \cite{Gubser:2008wv}
with two vector order parameters. Namely, a bulk model possessing two non-Abelian interacting vectors.

As it will emerge from the detailed analysis illustrated later, the two vectorial orderings arising in our model
seem a reasonable description of a p-wave and a ferromagnetic-like order parameters.
In this sense, the present holographic system describes the occurrence of p-wave superconductivity ``on the 
border'' of ferromagnetism in a strongly coupled context, where the mutual effects of ferromagnetism and 
superconductivity are equally included. More specifically, it features a concomitant transition from
a non-superconducting normal phase to a condensed phase with a p-wave ferromagnetic condensate.
This represents an extreme case of cooperation between two order parameters where 
each is necessary to the other and they present the same characteristic energy scale.

In the standard BCS theory, superconductivity emerges as a consequence of the formation of Cooper-pairs.
These pairs form because of a phonon-mediated attractive interaction between electrons with opposite spin.
Shortly after the formulation of the BCS theory, it was realized that spin-spin interaction could also mediate the coupling leading
to superconductivity \cite{Berk:1966} indicating that itinerant electron magnetism could be a novel resource for superconductivity \cite{Monthoux:2007}. 
This has been shown to be relevant in $^3$He \cite{Nakajima:73,Brinkman:74}, in heavy-fermions compounds \cite{Mathur:1998} and, finally, in HTC supercoductors \cite{bednorz:86}, 
where the superconductivity appears at the border of antiferromagnetism.  Spin-spin interactions can then have a great importance in the
panorama of superconducting systems. In this contest, an equal-spin-pairing superconducting state has been suggested in itinerant ferromagnets;
the mechanism involves a triplet p-wave order parameter where the electrons in the Cooper-pairs have equally oriented spins \cite{Fay:1980wv}. 
We then had to wait the new millennium to see, at the experimental level, some materials showing superconductivity at the border of itinerant 
ferromagnetism \cite{Saxena:2000,Pfleiderer:2001}. Nowadays there are many different triplet p-wave superconductors with strongly anysotropic 
ferromagnetism where the superconductivity is probably mediated by Ising ferromagnetic spin fluctuations \cite{Aoki:2001,Akazawa:2004,Huy:2007}. 
The peculiar behaviour of the critical field anisotropy has been investigated theoretically and it is evoked (at least for UCoGe) 
as an indication of the strong-coupling nature of the superconductivity \cite{Mineev2011,champel,Tada:11}. 

We have seen, that the holographic method has been proposed for the theoretical investigation of
HTCs. Our model can be read as a holographic viable way to generalize at strong coupling a p-wave superconducting 
mechanism analogous to that advanced in \cite{Fay:1980wv,Mineev2011,champel} which is concomitant with itinerant ferromagnetism.

In the last 30 years the numbers of strongly
correlated and unconventional superconductors has grown enormously \cite{Monthoux:2007}. Inspired by the layered
perovskite structure of some HTCs, the $\text{Sr}_2\text{RuO}_4$ was identified \cite{Maeno:1994}. In this material
the p-wave superconductivity is observed near the
antiferromagnetic phase.  At the same time the heavy fermion compounds, such as
$\text{CePd}_2\text{Si}_2$ and $\text{CeIn}_3$ \cite{Mathur:1998}
has shown a magnetically mediated superconductivity around the critical density of
the antiferromagnetic order and $\text{UPt}_3$ \cite{Joynt:2002}
presents multiple superconducting phases. A variant of the last compounds, the
$\text{UGe}_2$, has shown, under high pressure, a superconductivity transition on the
border of ferromagnetism \cite{Saxena:2000}. This
was interpreted as one of the first realizations of an itinerant ferromagnetic
superconductor  \cite{Fay:1980wv} with the material $\text{ZrZn}_2$ \cite{Pfleiderer:2001} as reported almost at the same time.
Since then, the number of ferromagnetic superconductors has grown with URhGe \cite{Aoki:2001} where ferromagnetic superconductivity was firstly
observed at ambient pressure, UIr \cite{Akazawa:2004} where inversion symmetry is broken and finally UCoGe \cite{Huy:2007}.

We think that, in our perspective, UCoGe is particularly interesting because it has a phase diagram in terms of the 
pressure which is completely different from that of other systems \cite{Aoki2011573}. More specifically, it has been
shown that at finite ambient pressure (around $1.2$ GPa) the Curie temperature
$T_{Cu}$ of the ferromagnetic transition coincides with the superconducting
transition $T_c$ \cite{Aoki2010}. In our model featuring concomitant condensation of the two order parameters, this is indeed the case. 
In addition, for this material, an unusual 
anisotropic behaviour for the critical fields has been shown; such feature suggests a strong coupling superconductivity \cite{Tada:11}.
Furthermore, we limit our analysis to 2+1 dimensions even though, also at the experimental level, 
it is still unclear whether in order to describe this system a 2+1 dimensions or a 3+1 dimensions model is more appropriate. 
Our main target is indeed to clarify how the interplay of the two order parameters influences the properties of the system. 
We think that the investigation of this point with
AdS/CFT correspondence is particularly useful because the two order
parameters has roughly the same energy scale and their interplay represents a
strongly coupled problem that cannot be simply approached with conventional methods
\cite{Fay:1980wv,Mineev2011,champel}.

\section{The model}
\label{model}

The bulk theory is a $3+1$ dimensional gravitational model with two interacting massless vector fields;
the Lagrangian density is composed by a standard Einstein-Hilbert gravitational part and 
by the terms describing the dynamics of the vector fields, namely
\begin{equation}
\label{action}
S=\frac{1}{2 \kappa_4^2} \int dx^4\, \sqrt{-g}\, \left[\left(\mathcal{R}+\frac{6}{L^2}\right) 
-\frac{1}{4 q^2} F_{\mu \nu}^aF^{a\, \mu \nu}
-\frac{1}{4 q^2} Y_{\mu \nu}^aY^{a\, \mu \nu}
+\frac{c}{2 q^2} F_{\mu \nu}^aY^{a\, \mu \nu} \right],
\end{equation}
where $\kappa_4$ is the four-dimensional Newton constant. The main peculiarity of the model relies in the presence of the last interaction 
term which mixes the dynamics of the two vector fields%
\footnote{Gravity mixes the dynamics of the two vector fields as well, however
we are here mainly concerned with the analysis of the bulk model in the so-called probe
approximation where the gravitational interactions are neglected, see Subsection \eqref{probe}
for further details.}. 
The strength of the mixing is controlled by the coupling constant $c$. 
Notice that the non-Abelian nature of the vector fields is crucial; 
indeed, an analogous Lagrangian for two Abelian vector fields
is always trivially diagonalizable and can be cast in a form where there is no $FY$ mixing term.
Another important consequence of the non-Abelian
character of the vector fields is that the gauge invariance of the $FY$ term
constrains the two vectors to be associated to the same gauge transformations%
\footnote{In the Abelian case, since the adjoint representation is trivial, 
we would have no constraint relating the gauge transformations of the two vector fields.}.
As in this paper we are mainly concerned with an $SU(2)$ non-Abelian gauge group, our model presents
a unique $SU(2)$ gauge invariance. 

The most general Lagrangian invariant under the following $SU(2)$ gauge transformation
\begin{eqnarray}\label{gausym1}
 \delta A_\mu^a &=& \partial_\mu \theta^a + f^{abc} A_\mu^b \theta^c \equiv (D_\mu^{(A)}\theta)^a\\ \label{gausym2}
 \delta B_\mu^a &=& \partial_\mu \theta^a + f^{abc} B_\mu^b \theta^c \equiv (D_\mu^{(B)}\theta)^a\ ,
\end{eqnarray}
where $\theta^a$ is the gauge parameter function, is
\begin{equation}\label{laginv}
 -\frac{1}{4\, q_1^2}\, F^{a\, \mu\nu} F^a_{\mu\nu}
 -\frac{1}{4\, q_2^2}\, Y^{a\, \mu\nu} Y^a_{\mu\nu}
 + c\; F^{a\, \mu\nu} Y^a_{\mu\nu}\ ,
\end{equation}
where $F^a_{\mu\nu}$ and $Y^a_{\mu\nu}$ are the field strengths for the 
two non-Abelian gauge fields $A_\mu^a$ and $B_\mu^a$ respectively,
\begin{equation}
\begin{split}
& F_{\mu \nu}^a=\partial_{\mu}A^a_{\nu}-\partial_{\nu} A^a_{\mu}+f^{abc}A_{\mu}^b A_{\nu}^c\\
& Y_{\mu \nu}^a=\partial_{\mu}B^a_{\nu}-\partial_{\nu} B^a_{\mu}+f^{abc}B_{\mu}^b B_{\nu}^c\ .
\end{split}
\end{equation}
In principle the Lagrangian density \eqref{laginv} depends on three generic coupling 
constants, namely $q_1^2$, $q_2^2$ and $c$.
The coupling constant $c$ is necessarily different from zero, because the 
mixing term $FY$ to which it is coupled is requested by the gauge symmetry
\eqref{gausym1} and \eqref{gausym2}. In other terms, the case $c=0$ is pathological and should be treated 
separately because, in that case, the gauge invariance is enhanced to 
$SU(2)\times SU(2)$. Such enhancement implies the presence of two gauge independent parameters (instead of one)
and two coupling constants (instead of three). 

In this paper we constrain the
two ``pure Yang-Mills'' coupling constants to be equal,
\begin{equation}
\label{charges}
 q_1^2 = q_2^2 = q^2\ .
\end{equation}
The resulting theory is then invariant under a single $SU(2)$ gauge symmetry
and depends on two coupling constants: $q^2$ and $c$.
Moreover, by rescaling the fields, it is possible to restrain 
ourselves to the case $q=1$ without losing generality.
The choice $q_1=q_2$ makes the action \eqref{action} invariant under the symmetry $A \leftrightarrow B$. 
However, the choice $q_1 \ne q_2$ could lead to interesting changes in the phase diagram of our model, 
and we postpone the discussion of such a case to future work \cite{progress}.

\subsection{The probe approximation and the gravitational background}
\label{probe}

The probe approximation consists in neglecting the back-reaction of the
vector fields on the geometry. At the level of the equations of motion,
it technically corresponds to considering the limit where the fields $A$ and $B$ are small \cite{Arias:2012py}. Indeed, in such a limit, 
the terms involving the vector fields appearing in the equations of motion
for the metric become negligible. 

We consider the following ansatz for the bulk metric
\begin{equation}\label{bulmet}
 ds^2 = - h(r)\, dt^2 + \frac{dr^2}{h(r)} + r^2 \left(dx^2 + dy^2 \right)\ .
\end{equation}
In the probe approximation the dynamics of the gravitational part of the 
model is decoupled from the rest and we are interested in considering 
an $AdS$-Schwarzschild black hole solution 
\begin{equation}\label{Schwarzschild}
 h(r) = \frac{r^2}{L^2} \left( 1 - \frac{r_H^3}{r^3}\right)\ , \ \ \ \
 \sqrt{-g(r)} = r^2\ ,
\end{equation}
where $r_H$ is the horizon radius and $L$ is the $AdS$ curvature radius.
The Hawking temperature of the black hole solution \eqref{Schwarzschild} is
\begin{equation}\label{hawtem}
T_H=\frac{3 }{4\pi } \frac{r_H}{L^2}\ .
\end{equation}
Since we are neglecting the backreaction of the gauge fields, these are regarded 
as small perturbations on the fixed black hole background which is therefore 
uncharged.

Once we have chosen the gravitational background, we can study the dynamics of the gauge fields.
From the action \eqref{action} we can derive the equations of motion for the
two vector fields obtaining
\begin{equation}
 g^{\rho\sigma} \mathbb{D}^{(A)}_{\rho} \left( \bm {F}_{\sigma\beta} - c\,  \bm{Y}_{\sigma\beta} \right)
= g^{\rho\sigma} \left[ D_\rho F^s_{\sigma\beta} + f^{sca} A^c_\rho F^a_{\sigma\beta} 
  -c \left(D_\rho Y^s_{\sigma\beta} + f^{sca} A^c_\rho Y^a_{\sigma\beta} \right)\right] = 0\ ,
\end{equation}
\begin{equation}
  g^{\rho\sigma} \mathbb{D}^{(B)}_{\rho} \left( \bm{Y}_{\sigma\beta} - c\,  \bm{F}_{\sigma\beta} \right)
= g^{\rho\sigma} \left[ D_\rho Y^s_{\sigma\beta} + f^{sca} B^c_\rho Y^a_{\sigma\beta} 
  -c \left(D_\rho F^s_{\sigma\beta} + f^{sca} B^c_\rho F^a_{\sigma\beta} \right)\right] = 0\ ,
\end{equation}
where $\mathbb{D}^{(X)}$ represents the covariant derivative both in relation to the gauge connection $X$
and to the metric connection.

\subsection{Definition of the physical quantities}
\label{physint}
In analogy with the analysis of the holographic p-wave superconductor 
\cite{Gubser:2008wv}, we consider the following ansatz
\begin{eqnarray}
 \label{ans_unp_A}
 \bm{A} &=& \Phi(r)\, \tau^3\, dt + W(r)\, \tau^1\, dx \ , \\
 \label{ans_unp_B}
 \bm{B} &=& H(r)\, \tau^3\, dt + V(r)\, \tau^1\, dx \ .
\end{eqnarray}
In general, an asymptotic study of the equations of motion unveils the large $r$ behaviour of the fields.
For the time being, let us concentrate on the temporal components whose asymptotic behaviour is 
\begin{equation}\label{asy}
 \Phi(r) = \mu_A - \frac{\rho_A}{r} + ...\ , \ \ \ \
 H(r) = \mu_B - \frac{\rho_B}{r} + ...\ . \ \ \ \
\end{equation}
Being $c \ne 0$, the $FY$ term in the action \eqref{action} mixes the kinetic terms of the 
two vector fields%
\footnote{\label{foot} Let us spend some words also about the $c=0$ case.
Here, before considering any particular ansatz, the fields $A_{\mu}^a$ and $B_{\mu}^a$ decouple from each others and
the gauge symmetry of the model is $SU(2) \times SU(2)$. 
In this case there are two conserved currents and, in accordance with the holographic dictionary \cite{Hartnoll:2009sz}, 
$\mu_A$, $\mu_B$, $\rho_A$ and $\rho_B$ are respectively the chemical potentials and the charge densities associated with the two sectors.}. 
Then the on-shell bulk action reduces to a boundary term of the following type
\begin{equation}
 \begin{split}\label{bou_unrot}
 &\left. \int_{\partial  {\cal M}}  d^3 x\ r^2 \left(A A' + c\, A B' + c\, B A' + B B'\right) \right|_{r=r_\infty} \\
 & \ \ \ \ \ \ \ \ \ \ \ \ \ \ \ \ \ \ \ \ \ \ \ \ \ \ \ \ \ \ \ \ \ \ \ 
 \sim V \left( \mu_A \rho_A + c\, \mu_A \rho_B + c\, \mu_B \rho_A + \mu_B \rho_B \right)\ ,
 \end{split}
\end{equation}
where $\partial {\cal M}$ represents the boundary $r \rightarrow \infty$ manifold and $V$ is its volume.
We have that $\mu_A$ and $\mu_B$ act as sources for some combination of $\rho_A$ and $\rho_B$.
In other terms, we lack a well-defined ``particle number'' in association to the two chemical species%
\footnote{A similar observation about diagonal chemical potentials and well-defined particle numbers
is mentioned in \cite{Krikun:2012yj} about a holographic model for a multi-band superconductor.}.

Let us observe that the terms in the left side of \eqref{bou_unrot} descend directly from the kinetic bulk terms. As we will see, 
their diagonalization corresponds, in the dual perspective, to define a well behaved chemical potential for the model. 
The diagonalization is achieved by means of the rotation:
\begin{eqnarray}\label{rotA}
 \bar{A}^{\, a}_\mu &=& \sqrt{\frac{1+c}{2}}\ (A^a_\mu - B^a_\mu)\\ \label{rotB}
 \bar{B}^{\, a}_\mu &=& \sqrt{\frac{1-c}{2}}\ (A^a_\mu + B^a_\mu)\ ,
\end{eqnarray}
whose inverse is 
\begin{eqnarray}\label{inv_rot}
  A^a_\mu &=& \frac{1}{\sqrt{2}} \left( \frac{1}{\sqrt{1+c}}\, \bar{A}^{\, a}_\mu + \frac{1}{\sqrt{1-c}}\, \bar{B}^{\, a}_\mu \right),\\
  \label{rot1}
  B^a_\mu &=& \frac{1}{\sqrt{2}} \left(-\frac{1}{\sqrt{1+c}}\, \bar{A}^{\, a}_\mu + \frac{1}{\sqrt{1-c}}\, \bar{B}^{\, a}_\mu \right)\ .
\end{eqnarray}
From the asymptotic behaviour of
$A_t$ and $B_t$ given in \eqref{asy}, we obtain an analogous large $r$ behavior for the 
rotated temporal components $\bar{A}_t$ and $\bar{B}_t$, namely
\begin{equation}\label{asy_rotated}
 \bar{A}^{\, 3}_t(r) = \mu_{\, \bar{A}} - \frac{\rho_{\, \bar{A}}}{r} + ...\ , \ \ \ \
 \bar{B}^{\, 3}_t(r) = \mu_{\, \bar{B}} - \frac{\rho_{\, \bar{B}}}{r} + ...\ .
\end{equation}
Eventually, after the rotation, the boundary term \eqref{bou_unrot} becomes
\begin{equation}
\label{acrot}
 \left. \int_{\partial {\cal M}} d^3 x\ r^2 \left(\bar{A}\; \bar{A}' + \bar{B}\; \bar{B}'\right) \right|_{r=r_\infty}
 \sim V \left( \mu_{\, \bar{A}}\; \rho_{\,\bar{A}}\; + \mu_{\,\bar{B}}\; \rho_{\,\bar{B}} \right)\ ,
\end{equation}

In relation to the complete model (i.e. before the introduction of any ansatz) the redefinitions \eqref{inv_rot} 
and \eqref{rot1} allow an interesting interpretation of the field 
content of the bulk theory which is hidden in our original action \eqref{action}.
The $SU(2)$ gauge transformations \eqref{gausym1} and \eqref{gausym2}, written in terms 
of the ``rotated'' field $\bar{A}_\mu^a$ and $\bar{B}_\mu^a$, read
\begin{eqnarray}\, \label{Bbar}
 \delta \bar{B}^a_\mu &=& \partial_\mu \theta^a + f^{abc} \bar{B}^b_\mu \theta^c\\ \label{Abar}
 \delta \bar{A}^a_\mu &=& f^{abc}\, \bar{A}^b_\mu \theta^c\ ,
\end{eqnarray}
from which it is clear that we are dealing with a true gauge connection $\bar{B}^a_\mu$
interacting with a vectorial matter field $\bar{A}^a_\mu$, in the adjoint representation of 
the gauge group, which is $SU(2)$ in the present case. 
As a consequence, in our model there is one conserved current which is related to the true bulk gauge 
field $\bar{B}_{\mu}^a$ and a non-conserved current related to the adjoint vector $\bar{A}_{\mu}^a$. 
The field transformations \eqref{rotA} and \eqref{rotB} are exactly those which make evident this peculiar 
symmetry structure, as outlined in \eqref{acrot}.

Since the source $\mu_{\bar{A}}$ is not associated to a bulk gauge field, in general it does not represent
an authentic chemical potential. Nevertheless, there is a subtle and important point regarding the role of the ansatz; 
as we will argue more in detail in Subsection \ref{exspo}, the normal phase of our model where only the temporal $\sigma^3$
components of the gauge fields are non-trivial, is characterized by a $U(1)\times U(1)$ symmetry where the two sectors transform independently.
This is related to the Abelian character of the $U(1)$ symmetry preserving $\sigma_3$. The ``Abelianization'' induced 
by the ansatz makes the two sectors independent also for $c\neq0$.
Therefore, similarly to the $c=0$ case (see footnote \ref{foot}), the normal phase shows two conserved currents and it is formally analogous
to the normal phase of the s-wave unbalanced holographic superconductor \cite{Bigazzi:2011ak,Musso:2013rva}.
For this reason, with a slight abuse, we keep the notation $\mu_{\bar{A}}$ throughout the whole paper also in relation
to the condensed phase even though there the $\bar{A}$ current is broken by the presence of the interaction terms
involving the condensed fields.

In the spirit of our phenomenological interpretation of the model as an effective theory which describes strongly correlated 
ferromagnetic superconductors, we interpret the conserved current associated to $\bar{B}_{\mu}^a$ as the charge density current, 
and the non-conserved current associated to $\bar{A}_{\mu}^a$ as the spin density current. This latter is not conserved in the condensed phase 
due to the interplay between the spin and the charge sectors.
In the condensed phase, $\mu_{\bar{B}}$ is a true chemical potential, while $\mu_{\bar{A}}$ is a source for the non-conserved 
spin density $\rho_{\bar{A}}$ which takes into account in an effective manner all the effects which contribute to a non vanishing 
spin magnetization in a strongly correlated ferromagnetic superconductor.\footnote{We thank the referee who suggested us to clarify this point.} 

Finally we make some comments on the value of the coupling $c$. At first we note that the asymptotic behaviour of the vector fields $A$ and $B$ 
does not depend on the particular value of $c$. Remarkably, due to this fact the current associated to $A$ has conformal dimension two also if 
$c \ne 0$, where is in general not conserved.
Secondly, observe that the coefficients of the rotation depend on the
coupling constant $c$ associated to the mixing and the rotation itself becomes
singular for the specific values $c=\pm 1$. Indeed, when $c=\pm 1$, 
the action of the two original unbarred vector fields reduces to 
the perfect square of $F\mp Y$. As a consequence, for these particular values
of $c$, only a combination of the two vector fields actually propagates. 
In the following we will avoid such limiting circumstances restraining 
ourselves to $c\neq\pm 1$.

\section{Analysis of the model in the ``physical'' basis}

From now on we adopt the rotated basis and refer to the following notation
\begin{eqnarray}
 \label{ans_phy_A}
 \bar{\bm{A}} &=& \phi(r) \tau^3 dt + w(r) \tau^1 dx \ ,\\
 \label{ans_phy_B}
 \bar{\bm{B}} &=& \eta(r) \tau^3 dt + v(r) \tau^1 dx \ .
\end{eqnarray}
Notice that the rotation does not affect the kind of ansatz we consider which 
has actually the same form before (see \eqref{ans_unp_A} and \eqref{ans_unp_B}) 
and after (see \eqref{ans_phy_A} and \eqref{ans_phy_B}) the field redefinitions. 
The large $r$ asymptotic behavior of the non-trivial field components are
\begin{equation}\label{asy_phi_eta}
 \phi(r) = \mu_\phi - \frac{\rho_\phi}{r} + ...\ , \ \ \ \
 \eta(r) = \mu_\eta - \frac{\rho_\eta}{r} + ...\ ,
\end{equation}
\begin{equation}\label{asy_w_v}
 w(r) = S_w - \frac{\mathcal{h} O_w \mathcal{i}}{r} + ...\ , \ \ \ \
 v(r) = S_v - \frac{\mathcal{h} O_v \mathcal{i}}{r} + ...\ .
\end{equation}
Note that \eqref{asy_phi_eta} is just a rewriting of \eqref{asy_rotated} with a different notation.
As we are interested in spontaneous (i.e. unsourced) condensations where the VEV's of 
the operators $O$ are non-trivial, the sources $S_w$ and $S_v$ are taken to be null.

\subsection{Physical fields}

Referring to the ansatz \eqref{ans_phy_A} and \eqref{ans_phy_B},
the Lagrangian density for the vector fields in \eqref{action} 
can be rewritten as follows
\begin{equation}\label{Lag_fis}
 \begin{split}
 \sqrt{g(r)}\, {\cal L} = & -\frac{1}{2} h(r) (w')^2
                            -\frac{1}{2} h(r) (v')^2
                            +\frac{1}{2} r^2 (\phi')^2
                            +\frac{1}{2} r^2 (\eta')^2 \\
                          &\ \ \ \ \frac{1}{4(1-c)h(r)} \left[v^2 \eta^2 
                                                            + w^2 \eta^2 
                                                            + \frac{4}{1+c} v w \eta \phi 
                                                            + v^2 \phi^2 + \frac{(1-c)^2}{(1+c)^2} w^2 \phi^2 \right]\ ,
 \end{split}
\end{equation}
where the primes indicate derivatives with respect to the radial coordinate $r$.

From the Lagrangian density \eqref{Lag_fis}, we can derive the equations of motion
\begin{equation}\label{eqphi}
 \phi'' + \frac{2}{r} \phi' - \frac{1}{2}\, \frac{1}{1-c}\, \frac{1}{r^2 h(r)}
 \left[\frac{2}{1+c} v w \eta + \phi v^2 + \frac{(1-c)^2}{(1+c)^2} \phi w^2\right] = 0
\end{equation}
\begin{equation}\label{eqw}
 w'' + \frac{h'(r)}{h(r)}w' + \frac{1}{2}\, \frac{1}{1-c}\, \frac{1}{h^2(r)}
 \left[w \eta^2 + \frac{2}{1+c} v \eta \phi + \frac{(1-c)^2}{(1+c)^2} w \phi^2\right] = 0
\end{equation}
\begin{equation}\label{eqeta}
  \eta'' + \frac{2}{r}\eta' - \frac{1}{2}\, \frac{1}{1-c}\, \frac{1}{r^2 h(r)}
 \left[ \eta v^2 + \eta w^2 + \frac{2}{1+c} w v \phi \right] = 0
\end{equation}
\begin{equation}\label{eqv}
 v'' + \frac{h'(r)}{h(r)}v' + \frac{1}{2}\, \frac{1}{1-c}\, \frac{1}{h^2(r)}
 \left[v \eta^2 + \frac{2}{1+c} w \eta \phi + v \phi^2\right] = 0
\end{equation}

The system of the equations of motion is invariant under the following set of scalings:
\begin{equation}\label{resca}
r \rightarrow b\, r\ , \ \ \ \
(t,x,y) \rightarrow \frac{1}{b}\, (t,x,y)\ , \ \ \ \
(\phi,\eta,w,v) \rightarrow b\; (\phi,\eta,w,v) , \ \ \ \
h \rightarrow b^2 h\ ,
\end{equation}
where $b$ is a generic positive real number. This scalings allow us to set $r_H=1$. Notice that the scalings \eqref{resca}
preserve the asymptotic $AdS$ character of the bulk metric \eqref{bulmet}
which corresponds to having $h(r) \sim r^2$ in the large $r$ region.
Recall that asymptotic $AdS$-ness is the bulk feature encoding 
the UV conformal fixed point of the boundary theory. Consequently, the 
scalings \eqref{resca} are the dual bulk manifestation of the ultraviolet 
scale invariance of the boundary theory.

Field configurations related by a rescaling \eqref{resca} are equivalent
and only scaling invariant quantities have a physical significance.
We define the physical temperature of the boundary theory normalizing the 
bulk black hole Hawking temperature \eqref{hawtem} as follows
\begin{equation}
 \tilde{T} = \frac{T_H}{\sqrt{\mu_\phi^2 + \mu_\eta^2}}\ .
\end{equation}
From now on, we set $L=1$, which amounts to a choice of unit length.

The normalized temperature $\tilde{T}$ is invariant under the generic transformation 
\eqref{resca} as the original temperature is related to an inverse Euclidean time period.
Another relevant and scaling invariant quantity is the ratio between the source $\mu_{\phi}$ and the chemical potential $\mu_{\eta}$: $\mu_\phi/\mu_\eta$. In addition, also the condensates $O_{w}$
and $O_v$, being related to the sub-leading $1/r$ term in the UV expansion of the corresponding 
bulk fields, need to be rescaled. We define
\begin{equation}
 \mathcal{h} \tilde{O}_{w/v} \mathcal{i} = \frac{ \mathcal{h}  O_{w/v} \mathcal{i}}{\sqrt{\mu_\phi^2 + \mu_\eta^2}}\ .
\end{equation}

\subsection{The ansatz and the symmetries of the different phases}

\subsubsection{Explicit and spontaneous breaking}
\label{exspo}

Whenever the bulk fields $w$ and $v$ are vanishing, we have null condensates $O_w$ and $O_v$
and the dual system is in the normal phase. The presence of non-zero chemical potential $\mu_{\eta}$ and a non-zero source $\mu_{\phi}$
however forces the fields $\phi$ and $\eta$ to acquire a non-trivial bulk profile also in the 
normal phase. From the bulk standpoint, $\mu_{\eta}$ and $\mu_{\phi}$ act as sources for the corresponding 
bulk fields which are ``coloured'' under $SU(2)$ and then break the original symmetry upon 
acquiring a non-trivial profile. 

Let us first consider the standard p-wave superconductor with a single gauge field. 
This can be obtained from the present generalized model upon considering a trivial $B$
field and still adopting the ansatz \eqref{ans_unp_A} for the field $A$. Being the
temporal component of $A$ directed along $\tau_3$, whenever it has a non-trivial bulk profile
it breaks the original $SU(2)$ down to the $U(1)$ which preserves $\tau_3$. This
$U(1)$ is a symmetry of the normal phase which in turn is spontaneously broken to $\mathbb{Z}_2$ when 
the superconducting $w$ condensation occurs.

In our generalized model we have an analogous framework. Firstly, as we will describe in 
more detail later, the condensations of the two fields $w$ and $v$ are constrained to occur
always together by the structure of the model itself. Said otherwise, we have no phases 
where only one condensate is non-trivial. Therefore, in complete generality, our model is 
either in the normal phase with no condensates at all or in a doubly condensed phase. In the latter case
all the original symmetry is almost completely broken (we remain with just a residual $\mathbb{Z}_2$); 
this is apparent from the shape of the ansatz
which presents fields along both the $\tau_3$ and the $\tau_1$ directions. In the normal phase,
instead, we have that $v$ and $w$ are vanishing while $\phi$ and $\eta$ are not. The original
$SU(2)$ symmetry is explicitly broken down to an Abelian symmetry preserving $\tau_3$.
Such a residual symmetry of the normal phase is actually $U(1)\times U(1)$, being each Abelian
factor related to $\phi$ and $\eta$ respectively.

Some further comments are in order. As we have already noted in introducing the model, the 
mixing term $FY$ in the action \eqref{action} yields different consequences depending on the
Abelian or non-Abelian character of the gauge invariance. In particular, when the gauge group
is Abelian, the action can be cast in a form where $FY$ disappears. When instead the gauge 
group is non-Abelian, the term $FY$ cannot be eliminated with a field rotation (as it contains 
non-linear interactions) and it constrains the gauge transformations of the two gauge fields 
to be the same. Our model is initially non-Abelian. Nevertheless,
introducing non-trivial field profiles along $\tau_3$ (as we do to describe the normal phase)
we somewhat ``Abelianize'' the model which then presents an effective $U(1)\times U(1)$ symmetry. 
The two Abelian factors appear as independent symmetries of the normal phase. This unconventional 
kind of explicit symmetry breaking is related to the choice of the ansatz and to the different qualitative 
behavior of our system for Abelian and non-Abelian gauge group. 

Let us note that the system of equations of motion (\ref{eqphi}-\ref{eqv}) is 
invariant if we flip the sign of both the condensate fields $w$ and $v$. As we will see in studying the 
phase diagram of the model (see Subsection \ref{pha}), this means that the doubly condensed phase where 
both $w$ and $v$ have non-trivial bulk profile preserves a residual $\mathbb{Z}_2$ symmetry out of the 
full $U(1)\times U(1)$ of the normal phase. 

As a final comment, observe that the normal phase of the system at hand is completely analogous
to that appearing in the s-wave unbalanced holographic superconductor \cite{Bigazzi:2011ak}.
We will describe the implications of this parallel in Subsection \ref{norma}.

\section{Equilibrium}

\subsection{Double condensation}
\label{double}

The model under analysis does not admit any phase where only one
of the two order parameters condenses alone. Hence, only two possibilities 
can occur: either the system is in the normal phase where both order 
parameters are vanishing or it is in a coexistence phase where both 
order parameters acquire a vacuum expectation value. This in turn implies 
that both condensations take place at the same critical temperature independently 
of the values we consider for the parameters of the model (e.g. the coupling $c$)
or the chemical potential $\mu_{\eta}$. As we will see in the following, there are nevertheless 
different condensed phases differing among themselves in relation to the relative sign 
of the two condensates.

The impossibility of having a single non-trivial condensate emerges 
directly from the bulk equations of motion (\ref{eqphi}-\ref{eqv}).
Indeed, suppose that we wanted to obtain a solution where only the 
order parameter associated to the component field $v$ is non-trivial while
$w$ vanishes; the equation of motion of the latter, \eqref{eqw}, would reduce to
\begin{equation}\label{uc}
 v\ \eta\ \phi = 0\ ,
\end{equation}
which constrains at least one further bulk field to be zero. If we consider either 
$\eta$ or $\phi$ to be null, the system reduces to a standard p-wave holographic 
superconductor \cite{Gubser:2008wv}%
\footnote{Note however that in doing so the system (\ref{eqphi}-\ref{eqv}) returns
a p-wave superconductor with $c$-dependent coefficients.}. If instead we take $v=0$ we are in the normal 
phase.

Recall that we consider always spontaneous condensations, namely, 
from the bulk perspective, we always require a vanishing source term at the boundary 
for the associated bulk fields. As a consequence, a component field $v$ or $w$ with vanishing VEV 
is forced to have a null radial profile. Actually, as the radial equation for the bulk 
fields associated to the condensates is second order, when both the UV leading terms 
are zero we retrieve therefore the trivial solution.

We are considering the system in the probe approximation, however the same conclusion about 
the ``unavoidable coexistence'' of the two orderings holds for the backreacted case as well.
Indeed, we can consider a vanishing $w$ and repeat the above argument starting from the 
backreacted equation for $w$ reaching again the same constraint equation \eqref{uc}. 
We do not report the backreacted equations since we will treat them elsewhere \cite{progress}.

From the above discussion emerged that the two orderings associated to the fields $w$ and $v$
are necessary each to the other. 
The physical outcome is that their dynamics is clearly intertwined and the model shows 
an extreme case of mutual cooperation between the two order parameters. One natural interpretation
could be that the two order parameters are related to the condensation of the same degrees 
of freedom which are charged under both $U(1)$ factors of the normal phase symmetry.
For instance, we could think of a Cooper-like mechanism where the total spin of the pairs is 
not vanishing and preferably oriented along a specific direction; in such a situation, 
the superconducting condensation would imply a concomitant spin polarization of the carriers. As a further observation,
note that this would clearly represent a case of itinerant magnetization.

\subsection{Suppression/enhancement of the orderings and role of \texorpdfstring{$c$}{}}

The crucial ingredient in our bulk model is the interaction $c$ which couples
the two sectors (associated respectively to the original field strengths $F$ and $Y$).
We want now to characterize how the equilibrium of the system and the features of the 
condensates are affected by the coupling $c$. 
\begin{figure}
\centering
\includegraphics[width=80mm]{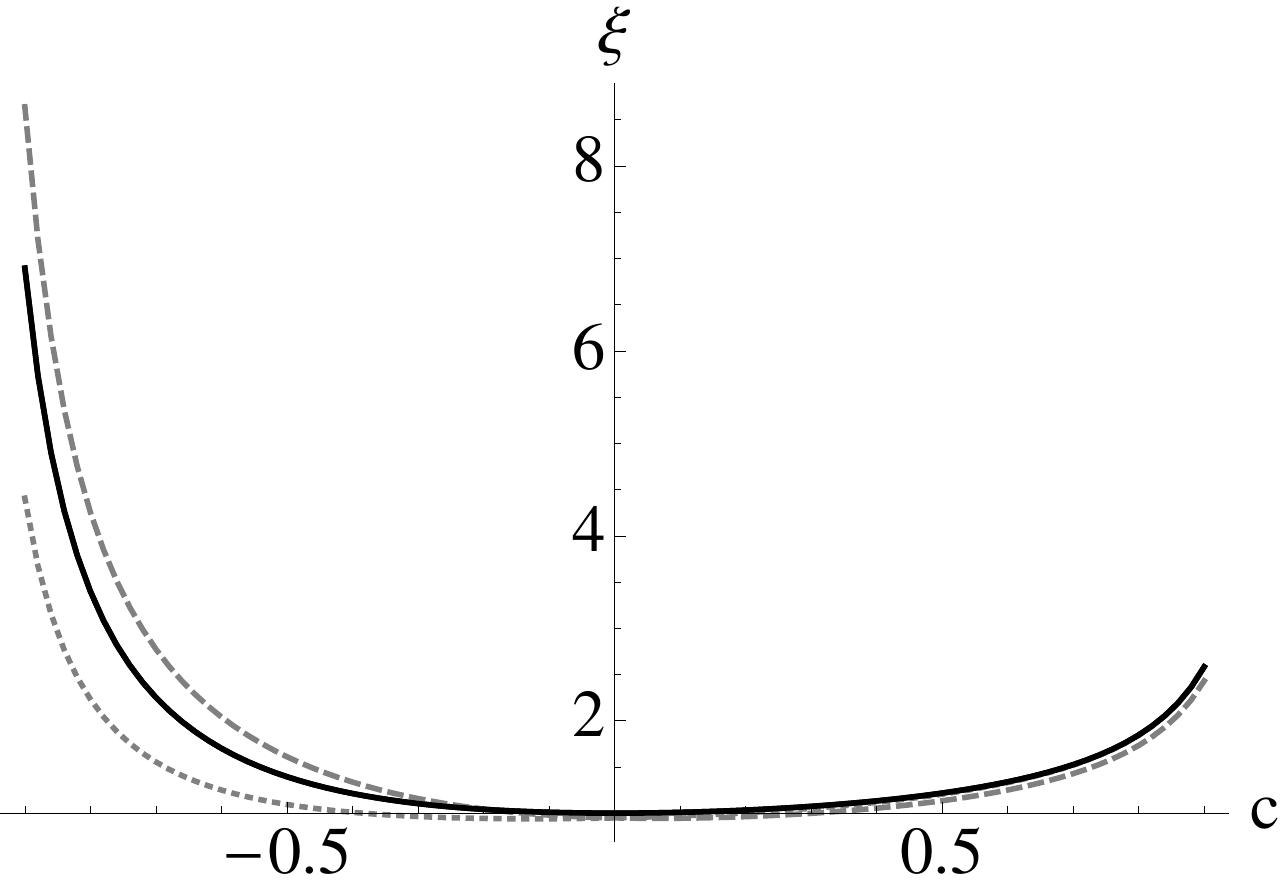} 
\caption{The condensation temperature as a function of the coupling $c$. 
The plot shows the quantity $\xi = \tilde{T}_c(c)/\tilde{T}^{(\mu_\phi=\mu_\eta)}_c(0)$ versus $c$. 
The solid line corresponds to $\mu_\phi = \mu_\eta$, the dashed line corresponds to 
$\mu_\phi / \mu_\eta = 2$ while the dotted line corresponds to $\mu_\phi / \mu_\eta = 1/2$.}
\label{Tcdic}
\end{figure}
In \cite{Basu:2010fa} the authors have proposed a conjecture stating that the 
enhancement/competition character of two orderings in the boundary theory 
corresponds respectively to a repulsive/attractive interaction of the 
associated bulk fields%
\footnote{The conjecture has been tested for a model of unbalanced superconductor
with two scalar order parameters in \cite{Musso:2013ija}}.
In the present system with two vectorial order parameters, the numerical results show that moving $c$
away from zero in either the positive or negative direction leads to a raise in the critical temperature
(see Figure \ref{Tcdic}).
This signals that the interaction yields an enhancement of the double condensation.
To rephrase the same concept, we have that increasing the absolute value of $c$
the condensed phase becomes thermodynamically favourable at a higher temperature.
Following the enhancement/repulsion conjecture proposed in \cite{Basu:2010fa},
we are lead to consider the effect of $c$ in the bulk as a repulsive interaction between 
the component fields $w$ and $v$%
\footnote{In the model with two scalar orderings \cite{Musso:2013ija},
the direct interaction between the order parameters are simple and their 
attractive/repulsive character is manifest already at the level of the Lagrangian.
Conversely, here the interaction terms between the order parameters are many.
So, in the main text, in line with \cite{Basu:2010fa}, we are lead to conjecture that their overall effect is always 
repulsive.}.

\begin{figure}
\centering
\includegraphics[width=75mm]{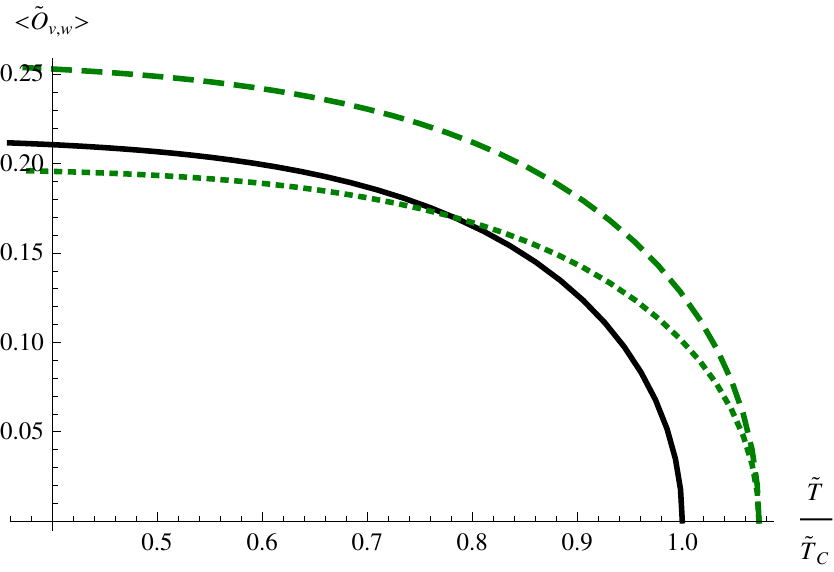} 
\includegraphics[width=75mm]{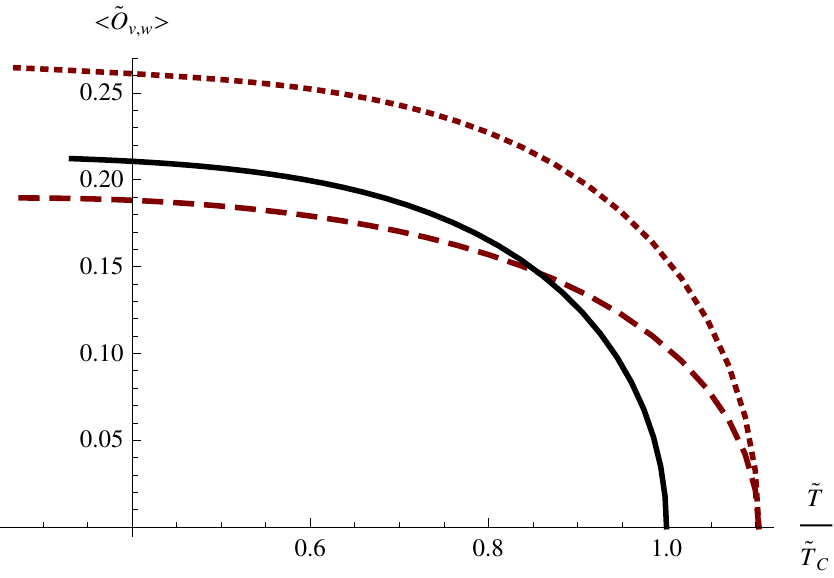} 
\caption{Order parameters $\mathcal{h} \mathcal{O}_v \mathcal{i}$ and $\mathcal{h} \mathcal{O}_w \mathcal{i}$ 
as a function of $T$ measured with the $T^{(c=0)}_c$ calculated for $c=0$. The black solid line refers to
the condensates obtained for $c=0$. In the left plot the two green lines are 
the condensates for $c=3/10$ while 
in the right plot the two red lines correspond to $c=-3/10$. In both plots the dashed
lines represent the $v$ condensate while the dotted lines represent the $w$ condensate. The plot were obtained with $\mu_{\phi} / \mu_{\eta}=1$}.
\label{condensates}
\end{figure}

Looking at the condensates from a closer perspective, we observe that the $c$ coupling has
further non-trivial effects. Indeed, even though we have insofar associated the 
enhancement or suppression effects to the position of the critical temperature only, one can
also wonder about the magnitude of the condensates. In this regard the system shows a fairly 
complicated behavior, see Figure \ref{condensates}. 
Depending on the sign of $c$, the interactions lead to a $v$ 
condensate always larger that the $w$ condensate for positive $c$. The opposite situation 
occurs for negative $c$. On top of that, we have that, although the condensation temperature
is always raised for bigger absolute values of $c$, the magnitude of the condensates can be suppressed 
with respect to the $c=0$ case for low enough temperature. More precisely, focus for instance at the 
left plot in Figure \ref{condensates}; there, for low enough temperature, the $w$ (dotted) condensate
becomes smaller when the coupling $c$ is turned on with respect to the $c=0$ case (solid line).
In this sense, when speaking about enhancement or suppression of an order parameter is important
to distinguish the effects on the critical temperature and the magnitude of the order parameters.

\subsection{Phase diagram}
\label{pha}

\begin{figure}
\centering
\includegraphics[width=90mm]{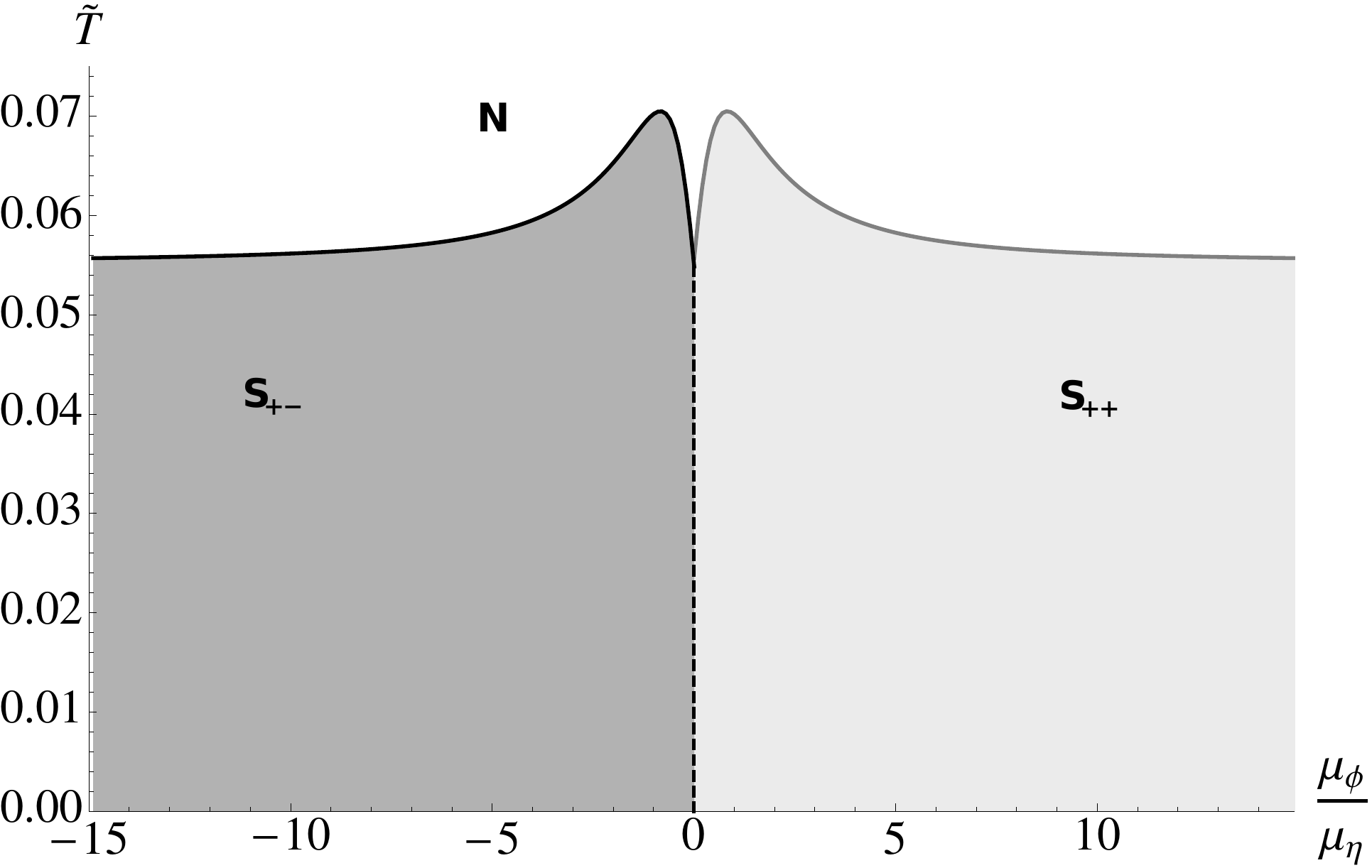} 
\caption{Phase diagram for $c=3/10$ on the plane $\tilde{T}$ versus $\mu_\phi/\mu_\eta$.
The high temperature white region corresponds to
the normal phase ({\itshape\sffamily {N}}). Lowering $T$ we can have to kinds of phases where the two condensates
(always condensing together) have either opposite sign (darker region, $\text{{\itshape\sffamily {S}}}_{{+-}}$) or the same sign 
(paler gray region, $\text{{\itshape\sffamily S}}_{++}$). Between the two condensed phases we have a first order transition (dashed line) while
the lines of transition connecting the condensed phases to the normal phase are second order (solid line).}
\label{phase_p3}
\end{figure}

As already noted in Subsection \ref{double}, with our choice $q_1=q_2$, our system can either be in the normal phase or in some doubly condensed phase.
In the latter, however, the condensates can have the same or opposite signs and this two cases correspond to two different phases%
\footnote{The concomitant change of sign to both condensates coincides with the 
already mentioned residual $\mathbb{Z}_2$ symmetry of the condensed phase.}. 
Figure \ref{phase_p3} contains the phase 
diagram obtained for $c=3/10$, though, for different values of the coupling $c$, the corresponding phase diagrams are 
qualitatively analogous and present the same overall phase structure (see Appendix \ref{phadif}). 
At high temperature the system is in the normal
phase ({\itshape\sffamily {N}})  while, at a sufficiently lower temperature, we always encounter a double condensation
for any value of the ratio $\mu_\phi/\mu_\eta$. 
More precisely, we find either condensed phase 
(i.e. with opposite $\text{{\itshape\sffamily S}}_{+-}$ or equal $\text{{\itshape\sffamily S}}_{++}$ signs) depending on the 
relative sign of the two chemical potentials $\mu_\phi$ and $\mu_\eta$.

To build the phase diagram contained in Figure \ref{phase_p3} it is necessary to study the
free energy of the system which allows us to single out the thermodynamically 
favoured phases for any point in the $\tilde{T}$ vs $\mu_\phi/\mu_\eta$ plane.
In line with the standard holographic dictionary, the free energy of the boundary theory 
is defined in terms of the Euclidean bulk action, namely
\begin{equation}\label{free}
 F =  -T\, \text{log}({\cal Z}) = -T\, S^{(\text{E})}_{\text{on-shell}}  
                          =  -T\, V\, \int_{r_H}^{+\infty} dr\, \sqrt{-g(r)}\, {\cal L}(r) \ ,                  
\end{equation}
where $V$ is the volume of the boundary theory manifold.
According to the scalings \eqref{resca}, the scaling dimension of the on-shell action is null.  Since the on-shell action is a scaling invariant quantity,
the free energy \eqref{free} scales as a temperature; to obtain the ``physical'' free energy
we have to normalize it in analogy with the temperature and consider
\begin{equation}
 \tilde{F} = - \tilde{T}\, S_{\text{on-shell}}\ .
\end{equation}

The normal phase ({\itshape\sffamily {N}}) is an analytically known solution of the bulk equations of motion
where the fields are given by
\begin{equation}\label{normafr}
 \phi^{(\text{{\itshape\sffamily {N}}})}(r) = \mu_\phi \left(1-\frac{r_H}{r}\right)\ , \ \ \ \ \
 \eta^{(\text{{\itshape\sffamily {N}}})}(r) = \mu_\eta \left(1-\frac{r_H}{r}\right)\ , \ \ \ \ \
 w^{(\text{{\itshape\sffamily {N}}})}(r) = 0\ , \ \ \ \ \
 v^{(\text{{\itshape\sffamily {N}}})}(r) = 0\ , 
\end{equation}
and the metric was already written in \eqref{Schwarzschild}.
Inserting \eqref{normafr} and \eqref{Schwarzschild} into the Lagrangian density \eqref{Lag_fis}, 
we obtain an analytical expression for the free energy in the normal phase,
\begin{equation}
 \tilde{F}^{({\text{{\itshape\sffamily {N}}}})} =  -\, \tilde{T}\, V\, \frac{r_H}{2}\, (\mu_\phi^2 + \mu_\eta^2)  \ .
\end{equation}
Since we do not have analytical control of the condensed phase, the computation of the 
corresponding free energy is done in terms of the numerical solutions.

\begin{figure}[ht]
\centering
\includegraphics[width=70mm]{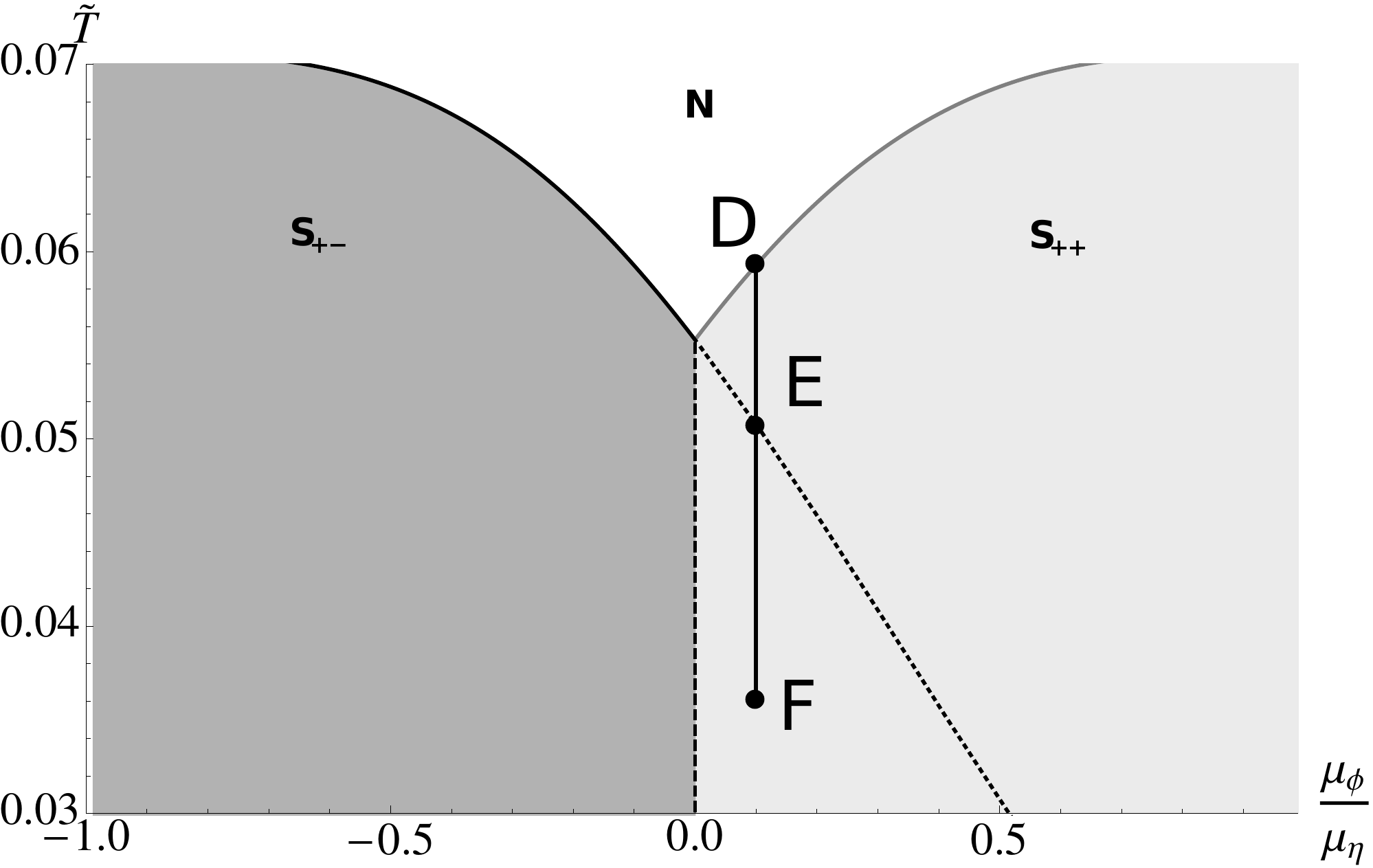} 
\includegraphics[width=70mm]{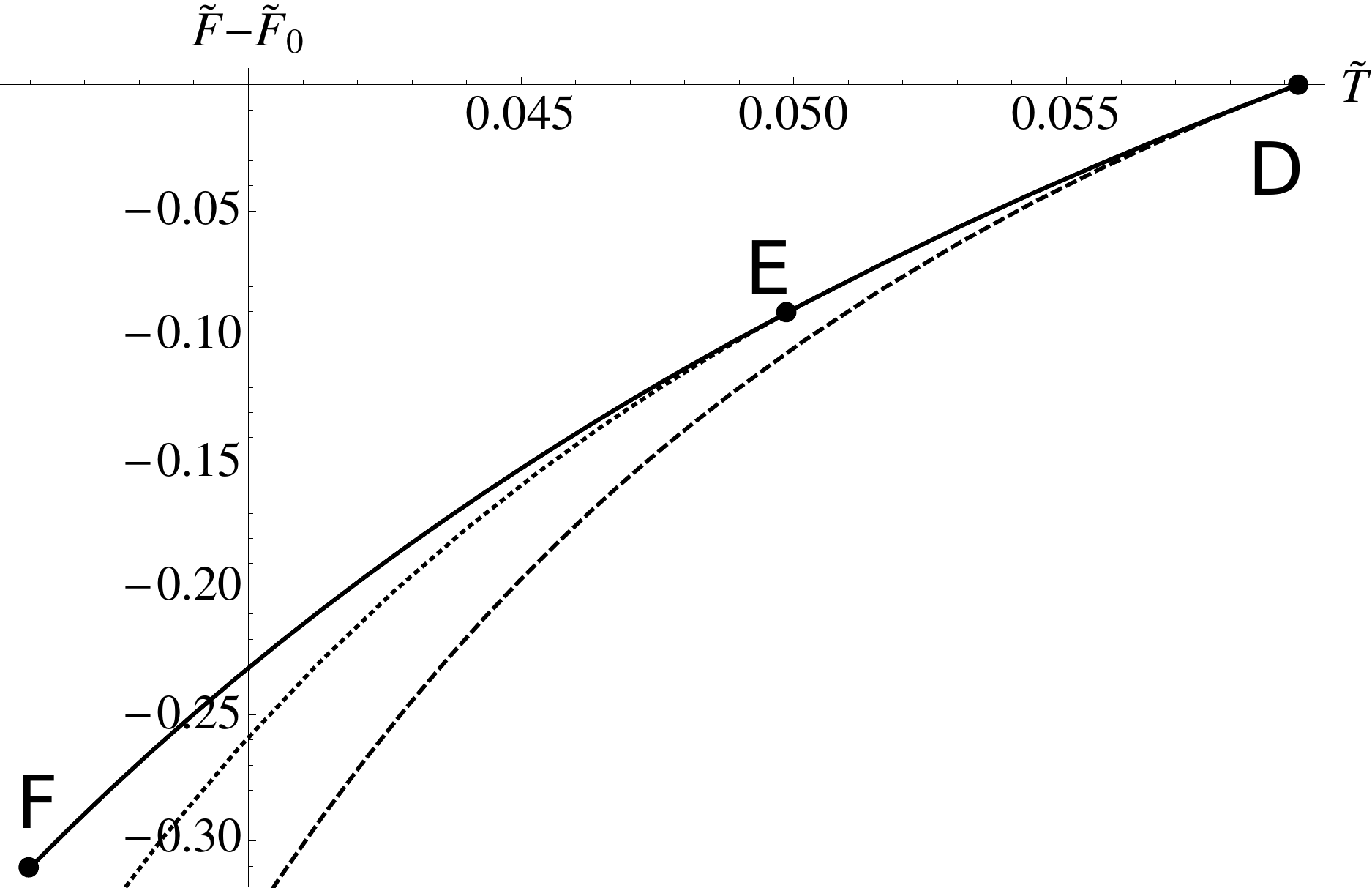} 
\caption{The second order transition between the normal and the condensed phase.
On the left plot we have traced a path at fixed $\mu_\phi/\mu_\eta$.
In the region below the dotted line two condensed phases (respectively with equal and opposite
signs for the condensates) are possible. In the right plot the solid line represents 
the free energy of the normal phase {\itshape \sffamily {N}}, the dashed line corresponds to the condensed phase with equal signs  $\text{{\itshape\sffamily S}}_{++}$
and the dotted line corresponds to the condensed phase with opposite signs  $\text{{\itshape\sffamily S}}_{+-}$.
We see that for $\mu_\phi/\mu_\eta$ greater than $0$ (i.e. when the chemical potential $\mu_{\eta}$ and the source $\mu_{\phi}$ have 
equal signs), the phase with equal signs for the condensates  $\text{{\itshape\sffamily S}}_{++}$
is always thermodynamically favoured. Observe also that the transition at $D$ is second order.
$\tilde{F}_0$ represents the value of the free energy at the transition.}
\label{free_energy}
\end{figure}
In Figure \ref{free_energy}
we plot the free energy of the various possible phases encountered while moving on a vertical path
in the phase diagram; namely a path obtained keeping the ratio $\mu_\phi/\mu_\eta$ fixed while lowering 
the temperature. The particular path traced in Figure \ref{free_energy} shows that, below $\tilde{T}_c$ and for 
$\mu_\phi/\mu_\eta>0$, the phase with equal signs for the condensates is always favoured.
This has actually lower free energy both with respect to the normal phase or the phase where the condensates
have opposite sign  $\text{{\itshape\sffamily S}}_{+-}$.
A symmetrical picture is obtained in the $\mu_\phi/\mu_\eta < 0$ half plane, where 
below $\tilde{T}_c$ we have always condensates with opposite sign.

\begin{figure}
\centering
\includegraphics[width=70mm]{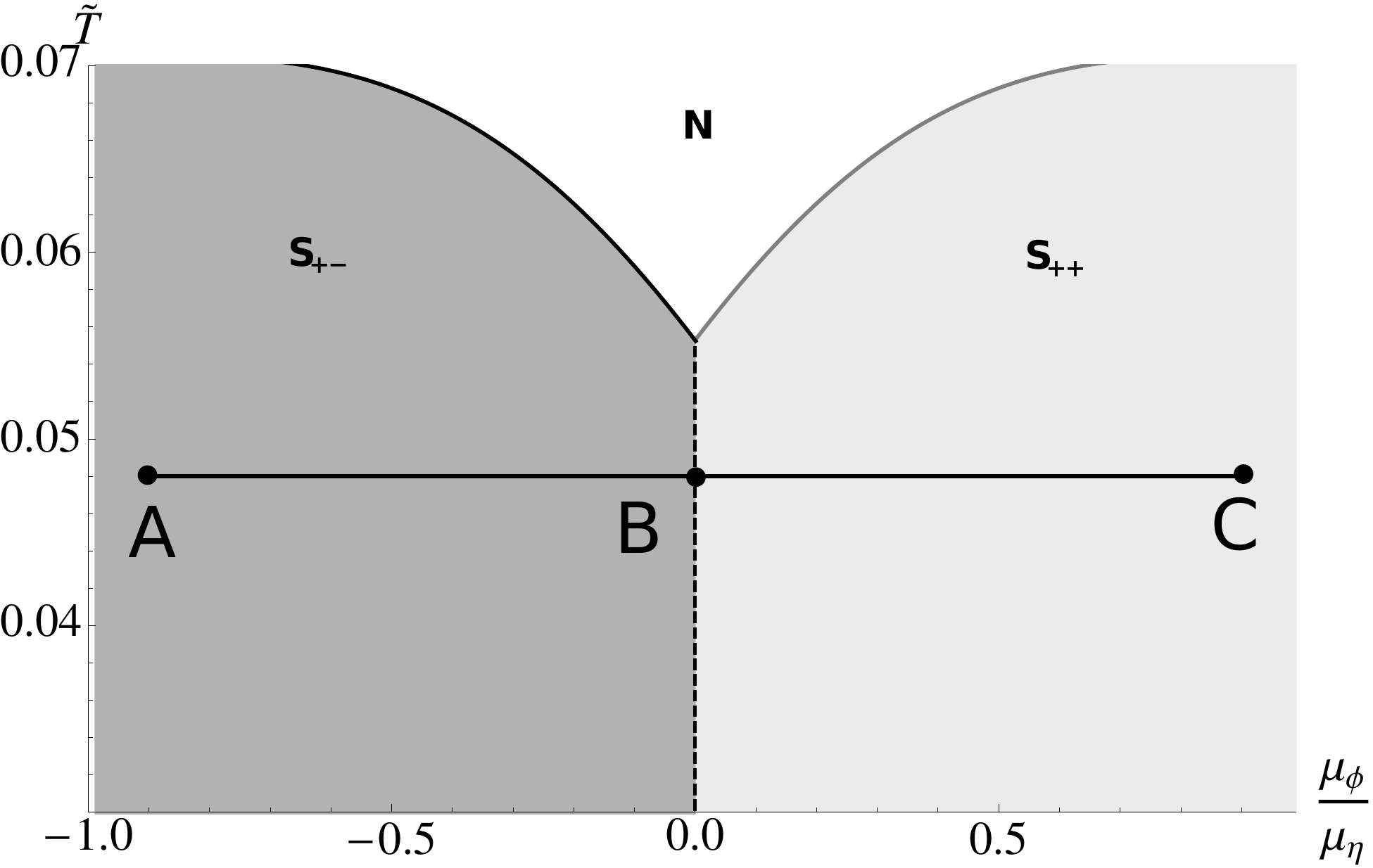} 
\includegraphics[width=70mm]{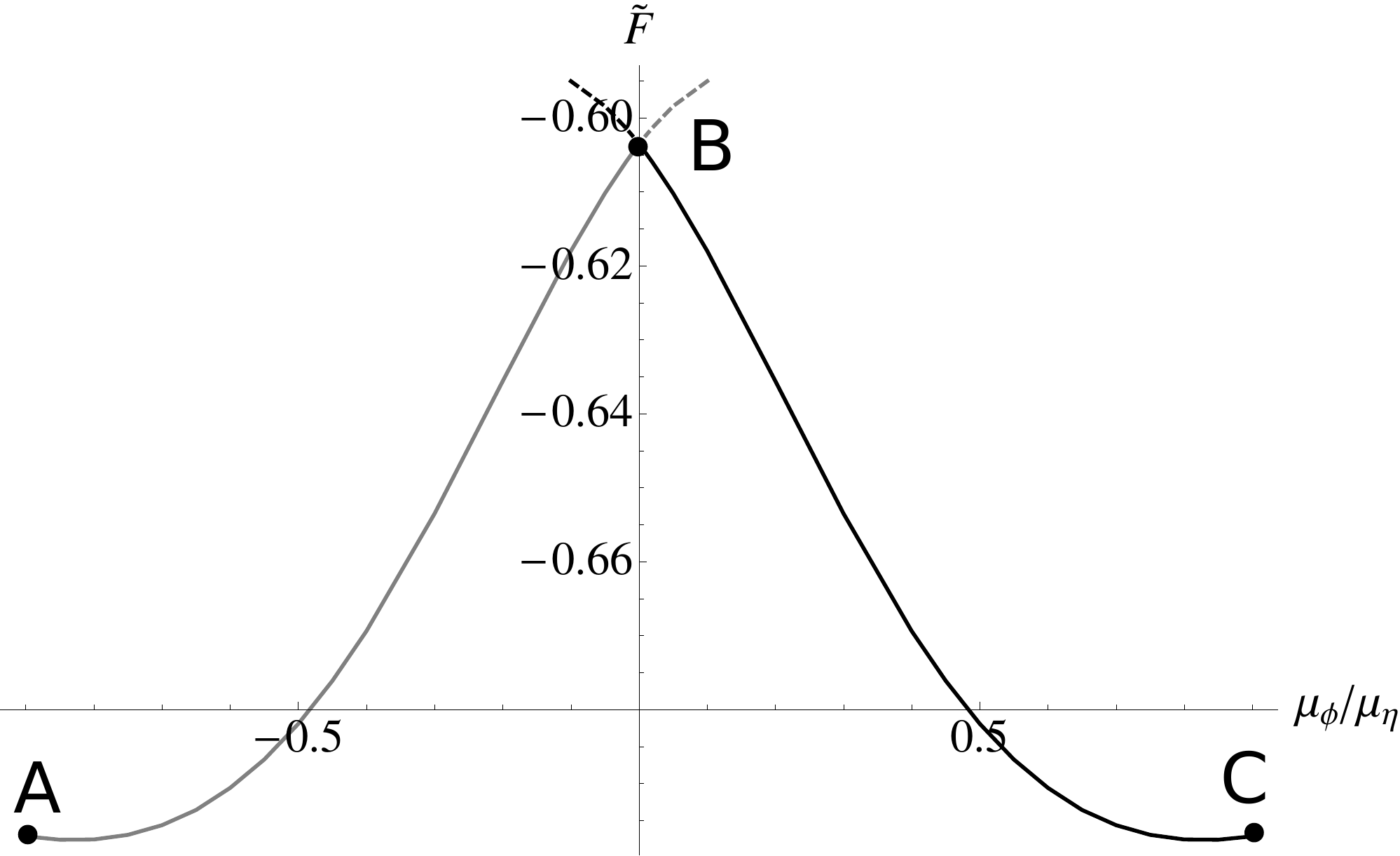} 
\caption{First order transition between the two condensed phases.
On the left plot, a fixed temperature path obtained moving $\mu_\phi/\mu_\eta$ is traced.
On the right plot it is shown the free energy along the path $ABC$. At $B$ there is a first order transition.
The dashed lines represent the extensions of the continued lines into a region where the associated 
phase is no longer thermodynamically favoured.}
\label{horizontal}
\end{figure}

The free energy of course allows us to study the character of the phase transitions which our 
system can undergo. Referring again to Figure \ref{free_energy}, we confirm that the transition 
between the normal phase and the condensed phase with equal signs is second order; this emerged 
already studying the shape of the condensates (see Figure \ref{condensates}) and it is corroborated by the observation that 
at the transition point (the point ${\bf D}$ in Figure \ref{free_energy}) the free energy plots
corresponding to the normal and condensed phases are tangent. Said otherwise, the discontinuity
in the free energy at the condensation is at least in the second derivative. Also in the 
$\mu_\phi/\mu_\eta < 0$ plane the picture is analogous and we have a second order transition between 
the normal and condensed phases.

Looking at the phase diagram \ref{phase_p3}, we see that another kind of transition is possible,
namely the one occurring when moving from one condensed phase to the other. Already the observation
that the two phases are distinguished by a different sign for one of the condensates, indicates that 
the transition is not continuous; note indeed that at a generic point close to the transition 
the condensates assume in general finite values. This guess can be precisely checked studying 
the behavior of the free energy at the transition (see Figure \ref{horizontal}). Actually,
at the transition point (point ${\bf B}$ in Figure \ref{horizontal}), we note that the free energy
presents a discontinuity already in its first derivative.

\section{Fluctuations around the bulk vacuum}
\label{tra}

In the previous sections we have shown that, at sufficiently high temperature, the vacuum 
solution of the bulk model possesses always a $U(1) \times U(1)$ 
gauge symmetry. From the dual perspective, this corresponds to a $U(1) \times U(1)$ global symmetry 
which characterizes the equilibrium state of the boundary theory in the normal phase. 
As we have already described, once the temperature is lowered, we always encounter a condensed phase which becomes energetically
favourable. Here the $U(1) \times U(1)$ symmetry is spontaneously broken down to $\mathbb{Z}_2$ by the 
concomitant condensations of two vectorial order parameters.

To study the linear response and transport properties of the boundary model, we have to focus on its 
slightly out-of-equilibrium features. On the bulk side, this corresponds to analyze
the small fluctuations of the system around the significant and stable vacuum; namely, the vacuum 
solution which is dual to a specified equilibrium state of the boundary theory we are interested in.

On a practical level, we first analyze the linearized equations of motion for the fluctuations of the bulk fields
focusing on the gauge invariant combinations; these are actually dual to the physical 
observables of the quantum field theory on the boundary. 
Next, we study the transport properties of the system both in its normal phase and
in the condensed, ferromagnetic/superconductor-like phase.

\subsection{Equations of motion}

We start our study of the fluctuations around a black hole configuration of the system by writing 
the full set of linearized equations for the bulk fluctuation fields.
We write the fields $A_{\mu}^a$ and $B_{\mu}^a$ as
\begin{eqnarray}
&A_{\mu}^a=\tilde{A}_{\mu}^a(r)+a_{\mu}^a(t,x,r)\ ,\\
&B_{\mu}^a=\tilde{B}_{\mu}^a(r)+b_{\mu}^a(t,x,r)\ ,
\end{eqnarray}
separating the fluctuation parts $a^a_\mu(t,x,r)$ and $b^a_\mu(t,x,r)$ from the background parts
$\tilde{A}_{\mu}^a(r)$ and $\tilde{B}_{\mu}^a(r)$. Notice that the background fields
coincide with the fields introduced in \eqref{ans_unp_A} and \eqref{ans_unp_B}; they 
arise from the study of the bulk solutions and encode the equilibrium of the boundary system. 
We impose the axial gauge fixing conditions $a_r^a=b_r^a=0$. 
We further assume that the fluctuations do not depend on the coordinate $y$. In relation to the other coordinates, we instead assume the following 
harmonic dependence:
\begin{eqnarray}
&a_{\mu}^a(t,x,r)=e^{-i\omega t+i \kappa x}\; a^a_{\mu}(r)\ ,\\
&b_{\mu}^a(t,x,r)=e^{-i\omega t+i \kappa x}\; b^a_{\mu}(r)\ .
\end{eqnarray}
In other words, we have plane waves in $t$ and $x$ while remaining generic along the radial coordinate $r$. 
As regards the coordinate $y$, our choice is equivalent to set the momentum along the $y$ direction to be null. 
We make this assumption since, as we shall see, we consider 
finite momentum only in the normal phase, which is isotropic, while in the condensed phase, for simplicity, we consider only null momentum.
We have chosen $a_{\mu}^a(t,x,r)$ and $b_{\mu}^a(t,x,r)$ to have the same momentum $\kappa$ and frequency 
$\omega$; as we shall see later, this choice is forced by the requirement of gauge invariance.

After considering the ``axial'' gauge choice $a_r=b_r=0$, there remain twelve equations of motion for the fields $a_i^a$ and 
$b_i^a$ (with $i={t,x}$ and $a$ labelling the $SU(2)$ adjoint representation) 
and six constraints; these latter arise from the equations of motion for 
the radial components $a_r^a$ and $b_r^a$. 
Due to the complexity of the equations, we prefer to write explicitly the whole system of differential
equations and the constraints only in Appendix \ref{lineeq}.

There are also six equations for the components $a_y^a$ and $b_y^a$; they describe
the transverse (with respect to the momentum) excitations of the model and, as the fluctuation fields do not depend on the coordinate $y$,
they decouple from the other fluctuations.
Also the system of equations of motion governing the transverse fluctuations along $y$ is reported
in Appendix \ref{lineeq}.

\subsection{Boundary conditions}

Since the equations for the fluctuations (see Appendix \ref{lineeq}) are of the Fuchsian type, 
we can use the term-wise Frobenius method and analyze the series expansions of the fields 
near the horizon and near the conformal boundary. 
For large values of the radial coordinate $r$ we obtain
\begin{eqnarray}
&a_{\mu}^a &= a_{\mu}^{a \; (\text{source})}+\frac{a_{\mu}^{a \; (\text{VEV})}}{r} + ... \ ,\\
&b_{\mu}^a &= b_{\mu}^{a \; (\text{source})}+\frac{b_{\mu}^{a \; (\text{VEV})}}{r} + ...\ ,
\end{eqnarray}
where, following the standard holographic prescription (see for instance \cite{Marolf:2006nd}),
the leading terms of the boundary expansions of the fields are interpreted as the sources 
for the dual operators in the boundary theory, while the subleading terms are interpreted 
as the VEV's of the same operators.

As we want to compute the retarded Green functions, we need to impose the ingoing wave boundary 
conditions at the horizon for the fluctuation fields $a_{\mu}^a$ and $b_{\mu}^a$
(see for instance \cite{Son:2002sd}), namely
\begin{eqnarray}
a_t^a&=&(1-r)^{1-\frac{i\omega}{3}}(a_t^{a \; (1)}+...),\\
b_t^a&=&(1-r)^{1-\frac{i\omega}{3}}(b_t^{a \; (1)}+...),\\
a_x^a&=&(1-r)^{-\frac{i\omega}{3}}(a_x^{a \;(0)}+...),\\
b_x^a&=&(1-r)^{-\frac{i\omega}{3}}(b_x^{a \;(0)}+...),\\
a_y^a&=&(1-r)^{-\frac{i\omega}{3}}(a_y^{a \;(0)}+...),\\
b_y^a&=&(1-r)^{-\frac{i\omega}{3}}(b_y^{a \;(0)}+...),
\end{eqnarray}
where $a_t^{a \; (1)}$, $b_t^{a \; (1)}$, $a_x^{a \; (0)}$, $b_x^{a \; (0)}$ 
$a_y^{a \; (0)}$ and $b_y^{a \; (0)}$ are arbitrary parameters.
As usual, in accordance with physical consistency (see \cite{Horowitz:2010gk}),
we have set to zero the leading term at the horizon of the temporal component 
of the vector fields $a_{t}^a$ and $b_t^a$  .

\subsection{Gauge invariant combinations of the fields}
\label{gaugeinvariant}

In this section we find the relevant gauge invariant combinations of the bulk fields which correspond
to the physical observables of the dual field theory at the boundary. 
Our analysis is similar to that illustrated in \cite{Gao:2012yw}.

\subsubsection{Gauge invariant combinations in the condensed phase} 
\label{brokenphase}

We start studying the gauge invariant combinations of the bulk fields in the broken phase. To this end,
we adopt the following method: at first we express the gauge symmetry of the fluctuations in the 
unphysical basis where the symmetry is easier to treat. 
Then, we use the gauge symmetry to derive the invariant combinations of the physical (barred) fields.
The physical fluctuations are again defined as the rotated background fields \eqref{rotA} and \eqref{rotB},
namely
\begin{eqnarray}
 & \bar{a}^a_{\mu} & =  \sqrt{\frac{1+c}{2}}\ (a_{\mu}^a-b_{\mu}^a) \ , \\
 & \bar{b}^a_{\mu} & = \sqrt{\frac{1+c}{2}}\ (a_{\mu}^a+b_{\mu}^a)\ .
\end{eqnarray}

As we have shown, our ansatz on the background fields \eqref{ans_unp_A}, \eqref{ans_unp_B} ``breaks 
explicitly'' the $SU(2)$ gauge invariance of the original action \eqref{action} down to $U(1) \times U(1)$. 
In the condensed phase, this symmetry is further spontaneously broken down to $\mathbb{Z}_2$ (associated to 
a global sign change of both condensates at the same time). 
Nonetheless, it is crucial to notice that the fluctuations around the vacuum with completely broken symmetry
(apart from the residual $\mathbb{Z}_2$) still transform under the full $SU(2)$ original symmetry, namely
\begin{eqnarray}
\label{sym11}
&\delta a_{\mu}^a(t,x,r)=\partial_{\mu} \alpha^a(t,x,r) + f^{abc} \tilde{A}_{\mu}^b(r)\; \alpha^c(t,x,r)\ ,\\
\label{sym22}
&\delta b_{\mu}^a(t,x,r)=\partial_{\mu} \alpha^a(t,x,r) + f^{abc} \tilde{B}_{\mu}^b(r)\; \alpha^c(t,x,r)\ .
\end{eqnarray}
Note that the gauge parameter functions $\alpha^a$ must be the same for both the $a$ and $b$ transformations;
this in order for the action \eqref{action} to be invariant under \eqref{sym11} and \eqref{sym22}.

Our purpose is now to find the gauge invariant linear combinations of the fluctuations of the physical fields $\bar{a}$ and $\bar{b}$. 
To this aim, we compute the pure gauge solution for the fields $a^a_{\mu}$ and $b^a_{\mu}$.
We assume the gauge parameter functions to have the same harmonic dependence on $t$ and $x$ as the fluctuation fields:
\begin{eqnarray}
\label{1}
&\delta(e^{-i \omega t+i \kappa x} a_{\mu}^a(r))=\partial_{\mu} (e^{-i \omega t+i \kappa x} \alpha^a) + e^{-i \omega t+i \kappa x} f^{abc} \tilde{A}_{\mu}^b(r)\; \alpha^c,\\
\label{2}
&\delta(e^{-i \omega t+i \kappa x} b_{\mu}^a(r))=\partial_{\mu} (e^{-i \omega t+i \kappa x} \alpha^a) + e^{-i \omega t+i \kappa x} f^{abc} \tilde{B}_{\mu}^b(r)\; \alpha^c,
\end{eqnarray}
where $\alpha^a$ cannot depend on $r$ as we have chosen the axial gauge $a^a_r=b^a_r=0$. 
As a consequence, $\alpha^a$ is a constant vector. The previous identities, relating the transformations
of $a$ and $b$ to the same gauge parameter functions $\alpha$, make it evident that, because of gauge invariance,
the fields $a$ and $b$ must have the same momentum $\kappa$ and 
the same frequency $\omega$.

Taking into account the transformations \eqref{1} and \eqref{2},
the purpose is now to find the relevant gauge invariant combinations of the bulk fields. 
Actually, it is possible to define several distinct gauge invariant combinations of the physical fields, 
and for the detail of the calculation we refer the reader to Appendix \ref{gaugeinv}.
Nonetheless, we henceforth focus the attention on only two of them as, after some phenomenology-inspired 
argument, they are the relevant combinations in view of the interpretation of the model as a ferromagnetic 
superconductor.

We consider the following two gauge invariant combinations: 
\begin{eqnarray}
\label{gaugeinvb1}
&\hat{\bar{a}}_x^3=\bar{a}_x^3+\frac{w}{\phi}\bar{a}_t^1\\
\label{gaugeinvb2}
&\hat{\bar{b}}_x^3=\bar{b}_x^3+\frac{\tilde{\kappa}}{\tilde{\omega}}\bar{b}_t^3+v\frac{\eta \bar{b}_t^1+i\tilde{\omega}\bar{b}_t^2}{\eta^2-\tilde{\omega}^2},
\end{eqnarray}
where $\tilde{\omega}= \sqrt{2(1-c)}\; \omega$, $\tilde{\kappa}= \sqrt{2(1-c)}\; \kappa$, and 
$\phi(r)$, $w(r)$, $\eta(r)$ and $v(r)$ are the physical background fields \eqref{ans_phy_A} and \eqref{ans_phy_B}.
Note that, in the limit of zero momentum, the combination $\hat{\bar{b}}_x^3$ coincides with that
considered in \cite{Gubser:2008wv} in relation to the analysis of the transport properties of the 
holographic p-wave superconductor.

We have defined the ``significant'' gauge invariant combination for $\hat{\bar{b}}_x^3$ 
proceeding as follows:
roughly speaking, we took the field $\bar{b}_x^3$ and added suitable terms in order to compensate 
its gauge variation. In doing so we required well-defined zero-momentum and zero-condensate limits, (see Appendix \ref{gaugeinv}). 
This constitutes actually a physical input which proves necessary to have sensible results in circumstances where the 
momentum vanishes or, as it occurs near criticality, where the condensates approach zero.
Considering such a criterion, the only remaining possibilities are actually the definitions \eqref{gaugeinvb1} and \eqref{gaugeinvb2}.
Notice that the combination \eqref{gaugeinvb2} has a vanishing denominator for $\tilde{\omega} = \eta$.
The structure of this denominator corresponds to the presence of a real-frequency mode in the system.
The pole arises when one considers fluctuations over a background with non-trivial $v$ along $\tau_3$. More comments on these modes are given later.

The same analysis in the transverse sector (that is along $y$) is much simpler.
From the gauge transformation of the fluctuation fields specified in 
\eqref{1} and \eqref{2} emerges that the fluctuations fields along $y$ are 
actually gauge invariant. This being a direct consequence of having chosen null momentum
along $y$ and of the fact that the background fields are trivial along $y$.

\subsubsection{Gauge invariant combinations in the normal phase} 
\label{normalphaset}

In the normal phase, as we have seen, the $SU(2)$ symmetry is partially broken by the background ansatz. As a consequence, the fluctuations 
in the normal phase transform as
\begin{equation}
\label{symmetrystructure}
\begin{split}
&\delta a_{\mu}^a= \partial_{\mu} \alpha^a+f^{abc} \left(A_{\mu}^b \alpha^c+a_{\mu}^b \Lambda^3 \delta^c_3 \right)\\
&\delta b_{\mu}^a= \partial_{\mu} \alpha^a+f^{abc} \left(B_{\mu}^b \alpha^c+b_{\mu}^b \Lambda^3 \delta^c_3 \right),
\end{split}
\end{equation}
where $\Lambda^3$ is the gauge parameter of the unbroken background symmetry%
\footnote{Notice that, because of the $FY$ mixing term, the gauge variation of the fluctuations 
constrains the two originally independent parameters of the background $U(1) \times U(1)$ symmetry to be equal.}. 
As a consequence, in the normal phase it is possible to define only two relevant gauge invariant combinations out 
of the physical fields, namely
\begin{eqnarray}
\label{gaugeinvn1}
&\bar{e}^3_{b}=\kappa \bar{b}^3_t+\omega \bar{b}^3_x,\\
\label{gaugeinvn2}
&\bar{e}^3_{a}=\kappa\bar{a}^3_t+\omega \bar{a}^3_x.
\end{eqnarray}

\section{Linear response: analytical and numerical results}

Having set the framework in which one can study the fluctuations of the bulk model,
we are now able to analyze the transport properties of the dual boundary system. 
At first we focus on the normal phase and, in particular, we show that
here the excitation modes are gap-less and degenerate.
Secondly, we explore the condensed phase and we 
analyze the transport coefficients. In the condensed case, the numerical treatment
becomes quite complicated and then we present only the analysis 
in the limit of zero momentum ($\kappa\rightarrow 0$) which is easier to treat. 

\subsection{Linear response in the normal phase}
\label{norma}
In the normal phase it is easy to see that the equations of motion for the fields 
$a_t^3$, $a_x^3$, $b_t^3$ and $b_x^3$ decouple from the other equations of motion, (see Appendix \ref{lineeq}). 
Consequently, we have a system of four equations and two constraints for these four fields. 
Moreover, the system of equations is diagonal and independent of the value of the coupling $c$, 
once expressed in terms of the physical fields $\bar{a}_t^3$, $\bar{a}_x^3$, 
$\bar{b}_t^3$ and $\bar{b}_x^3$\footnote{This is in accordance with the fact that, 
as we have shown in Section \ref{normalphaset}, the only two gauge invariant combinations of 
physical fields which we can construct are \eqref{gaugeinvn1} and \eqref{gaugeinvn2}}:
\begin{equation}
\begin{split}
&\bar{b}_t^{3 ''}+\frac{2}{r}\bar{b}_t^{3 '}-\frac{1}{r^2h}(\kappa^2\bar{b}_t^3+\kappa \omega \bar{b}_x^3)=0\ , \qquad \bar{a}_t^{3 ''}+\frac{2}{r}\bar{a}_t^{3 '}-\frac{1}{r^2h}(\kappa^2\bar{a}_t^3+\kappa \omega \bar{a}_x^3)=0\ ,\\
&\bar{b}_x^{3 ''}+\frac{h'}{h}\bar{b}_x^{3 '}+\frac{1}{h^2}(\omega^2 \bar{b}_x^3 + \kappa \omega \bar{b}_t^3)=0\ , \qquad \bar{a}_x^{3 ''}+\frac{h'}{h}\bar{a}_x^{3 '}+\frac{1}{h^2}(\omega^2 \bar{a}_x^3 + \kappa \omega \bar{a}_t^3)=0\ ,\\
&r^2\omega \bar{b}_t^{3'}+ \kappa h \bar{b}_x^{3'}=0\ , \qquad \qquad \qquad \; \; \; \qquad r^2\omega \bar{a}_t^{3'}+ \kappa h \bar{a}_x^{3'}=0\ .
\end{split}
\end{equation} 
This system of equations, with the symmetry structure \eqref{symmetrystructure}, is identical to that studied 
in \cite{Miranda:2008vb,Kovtun:2005ev, Amado:2009ts}, 
where it was found that, in the hydrodynamic limit $\omega, \kappa\ll 1$, the poles of the holographic Green functions 
$\mathcal{h} \bar{e}^3_a \bar{e}^3_a \mathcal{i} \propto \frac{\bar{e}_a^{3'}}{\bar{e}_a^3}$ 
and $\mathcal{h} \bar{e}^3_b \bar{e}^3_b \mathcal{i} \propto \frac{\bar{e}_b^{3'}}{\bar{e}_b^3}$ lie on:
\begin{equation}\label{omega_kappa}
\omega= -i \kappa^2\ .
\end{equation}
This shows that, in the normal phase, the modes of both the $\bar A$ and the $\bar B$ field manifest an analogous diffusive behaviour. 
This happens due to the fact that in the normal phase the interaction terms between the charge sector associated with $\bar B$ 
and the spin sector associated with $\bar A$ (which, as we have explained in Section \ref{physint}, causes the non-conservation 
of the spin density current) vanish.

As we have already noticed, the normal phase of the system presents a strong analogy with 
the normal phase of the unbalanced holographic superconductor \cite{Bigazzi:2011ak}.
Indeed, when the condensates are trivial, only the temporal $\tau_3$ components are relevant 
and the model is effectively Abelian. Such effective Abelianization leads to the double $U(1)\times U(1)$
symmetry of the normal phase and to the insensitivity of the model to the coupling constant $c$. 
Moreover, the two gauge fields and their dynamics are identical to that of the two 
Abelian gauge fields of the unbalanced superconductor. The analogy is particularly interesting because it gives us 
for free important information about the backreacted transport properties of the ``double'' p-wave model
at hand. Specifically, we learn that the conductivity matrix can be parametrized in terms of a single 
frequency-dependent ``mobility'' function%
\footnote{To have more details see \cite{Bigazzi:2011ak,Musso:2012sn}}. 
This fact highlights that, already in the normal phase, the 
electric and spin transport properties are intimately intertwined once the backreaction is considered.

\subsection{Linear response in the condensed phase}
\label{traconde}
\paragraph{The longitudinal sector} $\;$\\
In the condensed phase at vanishing momentum the equations of motion for the fields 
$a_t^1$, $a_t^2$, $a_x^3$, $b_t^1$, $b_t^2$ and $b_x^3$ decouple from the others\footnote{For this reason 
we have particularly emphasized the gauge invariant combinations \eqref{gaugeinvb1} and \eqref{gaugeinvb2} 
in Section \ref{brokenphase}.} (see Appendix \ref{lineeq}). Therefore, we have a system of six equations
of motion and four constraints for the fields (in the unphysical basis), namely
\begin{equation}
 \label{siss1}
a_t^{1''}-c\,
   b_t^{1''}+\frac{2 \left(a_t^{1'}-c\,
   b_t^{1'}\right)}{r}-\frac{a_x^3 (c H 
   V-W \Phi )}{h\, r^2}=0\ ,\\
\end{equation}
\begin{equation}
a_t^{2''}-c\,
   b_t^{2''}+\frac{2
   \left(a_t^{2'}-c\,
   b_t^{2'}\right)}{r}-\frac{i W \omega  \left(a_x^3-c\, b_x^3\right)+W^2
   a_t^2-c\, V W b_t^2}{h r^2}=0\ ,
\end{equation}
\begin{equation}
 \begin{split}
a_x^{3''} & -c\,
   b_x^{3''}+\frac{h' \left(a_x^{3'}-c\,
   b_x^{3'}\right)}{h} +
   \\ &+ \frac{i \omega  \left(c V b_t^2-W a_t^2\right)+\omega ^2
   \left(a_x^3-c\, b_x^3\right)+c H  V a_t^1-W \Phi 
   a_t^1}{h^2}=0\ ,
 \end{split}
\end{equation}
\begin{equation}
\label{siss3}
-c\, r^2 H ' a_t^2-i r^2 \omega\,  a_t^{1'}+r^2
   \Phi ' a_t^2-r^2 \Phi\,  a_t^{2'}+i\, c\, r^2
   \omega\,  b_t^{1'}+c\, r^2 \Phi \,
   b_t^{2'}=0\ ,
\end{equation}
\begin{equation}
\label{siss2}
 \begin{split}
c\, h\, V' a_x^3 & +c\, r^2 H ' a_t^1-h W' a_x^3+h\, W
   a_x^{3'}-i r^2 \omega\, 
   a_t^{2'} \\ &  -r^2 \Phi ' a_t^1 +r^2 \Phi \,
  a_t^{1'}
   -c\, h\, W\, b_x^{3'}+i\, c\, r^2
   \omega\,  b_t^{2'}-c\, r^2 \Phi \,
   b_t^{1'}=0\ ,
 \end{split}  
\end{equation}
and, as illustrated in Appendix \ref{lineeq}, the other three equations and two constraints are 
obtained from the previous ones exchanging $a_{\mu}^a \leftrightarrow b_{\mu}^a$, $W \leftrightarrow V$ and $H \leftrightarrow \Phi$.

We focus our attention on this system since, once the value of $\omega$ is fixed, 
the solution of \eqref{siss1}-\eqref{siss2} is determined.
Let us concentrate on the independent degrees of freedom of the system 
\eqref{siss1}-\eqref{siss2}. We have six second order linear differential equations which then require twelve integration constants. 
Choosing the ingoing wave boundary conditions (in accordance to the holographic prescription to compute the dual 
retarded Green functions) the free parameters get halved, so we remain with $6$ of them.
Furthermore, setting to zero the leading term of the horizon expansion of $a_t^1,\; b_t^1,\;a_t^2,$ and $b_t^2$  
fixes $4$ additional parameters leaving us with only $2$ integration constants.
Having set the previous conditions, two of the four constraints \eqref{siss3}-\eqref{siss2} are automatically satisfied, 
while the other two provide the same relation between the leading term of $a_x^3$ and $b_x^3$ in the horizon expansion, 
and consequently reduce this two remaining degrees of freedom to one.
Eventually, the solution is unique up
to an overall scaling which allows us to set the value of $a_x^3$ at the horizon to one. 
So, as we have claimed, once we have fixed the value of $\omega$, the solution of the system \eqref{siss1}-\eqref{siss2} is completely 
specified.
\paragraph{The transverse sector} $\;$\\
In the transverse sector at vanishing momentum the equations of motion for $a_y^3$ and $b_y^3$ (in the unphysical basis) decouple 
from the other equations (see Appendix \ref{lineeq}):
\begin{equation}
\label{trasv1}
b_y^{3''}-c\, a_y^{3''}+\frac{h'
   \left(b_y^{3'}-c\,
   a_y^{3'}\right)}{h}+\frac{\omega ^2 \left(b_y^3-c\, a_y^3\right)}{h^2}+
\frac{-V^2
   b_y^3+c V W a_y^3}{h
   r^2}=0\ ,
\end{equation}
\begin{equation}
\label{trasv2}
a_y^{3''}-c\, b_y^{3''}+\frac{h'
   \left(a_y^{3'}-c\,
   b_y^{3'}\right)}{h}+\frac{\omega ^2 \left(a_y^3-c\, b_y^3\right)}{h^2}+
\frac{-W^2
   a_y^3+c V W b_y^3}{h
   r^2}=0\ .
\end{equation}
The transverse sector is then described by an unconstrained system of two second order 
linear differential equations whose generic solution belongs to a $4$-parameter family. 
Restraining ourselves to ingoing solutions, we fix two parameters. The fields $a^3_y$
and $b^3_y$ can be rescaled together mapping a solution into a rescaled solution. Fixing such
scaling symmetry reduces the number of free parameters to one.
The transverse system is different from the longitudinal system in that we do not have 
constraints which fix this last $1$-parameter freedom. 

Given the structure of the system of transverse equations, in order to compute the transverse 
conductivity, we need to disentangle the dependence of the current on the 
two independent sources. These latter being associated with the near-boundary leading terms
of the gauge invariant combinations $\bar{a}_y^3=\sqrt{\frac{1+c}{2}}\left( a_y^3-b_y^3 \right)$ and $\bar{b}_y^3=\sqrt{\frac{1-c}{2}}\left( a_y^3+b_y^3 \right)$.
On a practical level, the computations are in spirit completely analogous to those
presented in \cite{Bigazzi:2011ak} about the mixed spin/electric linear response of the unbalanced superconductor where
it was necessary to distinguish between the effects of an ``electric'' or ``magnetic'' source.
According to the linear response framework, we assume the transverse current to depend linearly 
on both sources in the following way:
\begin{equation}
\label{transportmatrix}
\begin{pmatrix} J_{\bar{a}}^y (\omega) \\ J_{\bar{b}}^y (\omega)  \end{pmatrix}= \begin{pmatrix} \chi_{ \bar{a} \bar{a}}(\omega) & \sigma_{\bar{a} \bar{b}}(\omega)\\ \sigma_{\bar{b} \bar{a}}(\omega) & \sigma_{\bar{b} \bar{b}}(\omega) \end{pmatrix} \begin{pmatrix}  \bar{a}^{y \; 3} (\omega) \\ \bar{b}^{y \; 3} (\omega) \end{pmatrix}.
\end{equation}
To obtain the purely electric and purely spin response we are interested in the corresponding 
diagonal entry of the transport matrix \eqref{transportmatrix}. We have then a linear matrix equation and the $1$-parameter 
freedom allows us to consider as many solutions of the differential system as we need to invert 
the linear response equation and determine the 
purely electric conductivity and purely spin susceptibility, as we shall explain in the following sections. 
For further details we refer to \cite{Bigazzi:2011ak}.

As a final comment, note that the transverse sector differs from the longitudinal one as
this latter is more constrained. In the longitudinal sector, the consistency with the constraints 
does not allow one to have two independent sources for the fluctuations. The analysis of the longitudinal 
response is accordingly simplified and such a constrained dynamics 
underlines once more the intimately intertwined character of the $a$ and $b$ longitudinal sectors.

\subsubsection{The conductivity} 

In the previous sections we have argued how the gauge symmetry 
allows us to interpret the order parameter associated with 
the vector field $\bar{B}_{\mu}^a$ as a superconducting order parameter. 
By virtue of this, we interpret its fluctuations 
around the vacuum as electric perturbations of the ground state of the theory at the boundary.
In particular, we are interested in computing the conductivity both in the $xx$ and in the $yy$ direction in order to characterize the eventual
anisotropic structure of the superconducting-like energy gap, as could be expected for the p-wave holographic superconductor \cite{Gubser:2008wv}. 
Notice that $xx$ is the same direction in which the spin density order parameter is aligned.
Recall that, to give an inaccurate but intuitive picture, such spin density order parameter 
can be thought as associated to the spins of the Cooper-like pairs forming the condensate%
\footnote{This is the relevant case in the mechanism of itinerant ferromagnetic superconductor introduced by \cite{Fay:1980wv}}.
The two order parameters are aligned by construction, namely, as a consequence 
of our ansatz. We have also considered ansatzes where the two bulk components associated 
to the order parameters are aligned along different directions; in such cases the equations 
of motion imply strict constraints and actually only trivial solutions are allowed%
\footnote{The same mixed ansatz (i.e. where the condensates are along different spatial and/or
``color'' directions) have been studied for larger non-Abelian groups as well. The outcome is that,
besides the situation where the condensates are aligned along the same color and spatial direction,
the system of equations of motion is too restrictive to yield non-trivial dynamics.}.
\paragraph{The longitudinal sector} $\;$\\
In line with the analysis performed in Section \ref{gaugeinvariant}, it is natural
to define the electric conductivity $\sigma_{xx}$ in relation to the gauge invariant 
combination of bulk fields \eqref{gaugeinvb2}. This is the combination shared by the present model and
the standard holographic p-wave superconductor \cite{Gubser:2008wv}.
We remind the reader that there are several gauge 
invariant combinations of the bulk fields $\bar{b}_{\mu}^a$ characterized by being aligned along
the $x$ direction. Nevertheless, it is important to underline that the gauge invariant combination \eqref{gaugeinvb2}
which we chose is the only one that is well defined in the limit of zero spatial momentum 
and in the probe approximation (i.e. near the phase transition where the condensates vanish).

As illustrated in the previous sections, the system of differential equations 
which is necessary to solve in order to compute the gauge invariant combination \eqref{gaugeinvb2} 
has a unique solution, and this fact facilitate our physical interpretation.
\begin{figure}[ht]
\centering
\includegraphics[width=75mm]{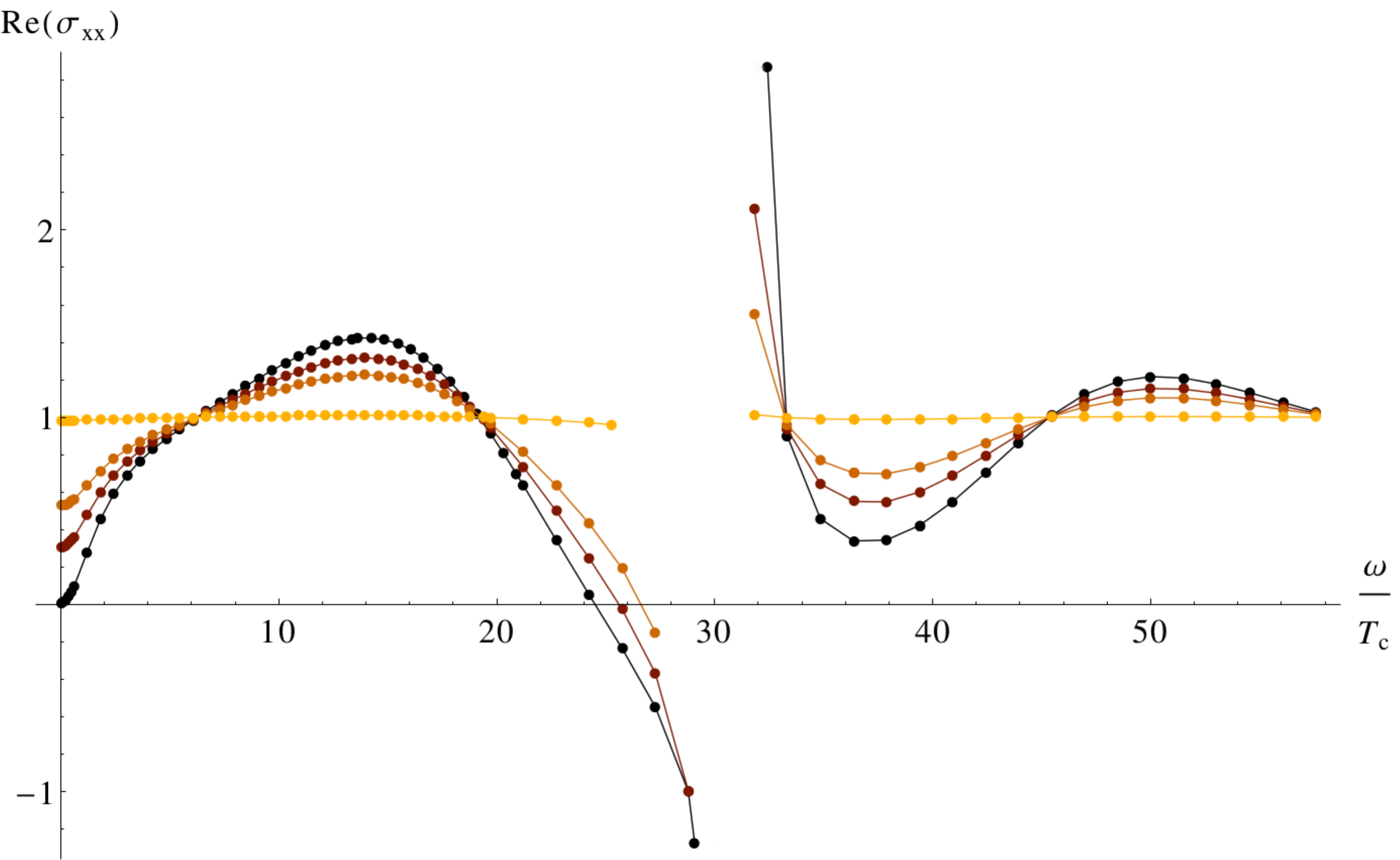} 
\includegraphics[width=75mm]{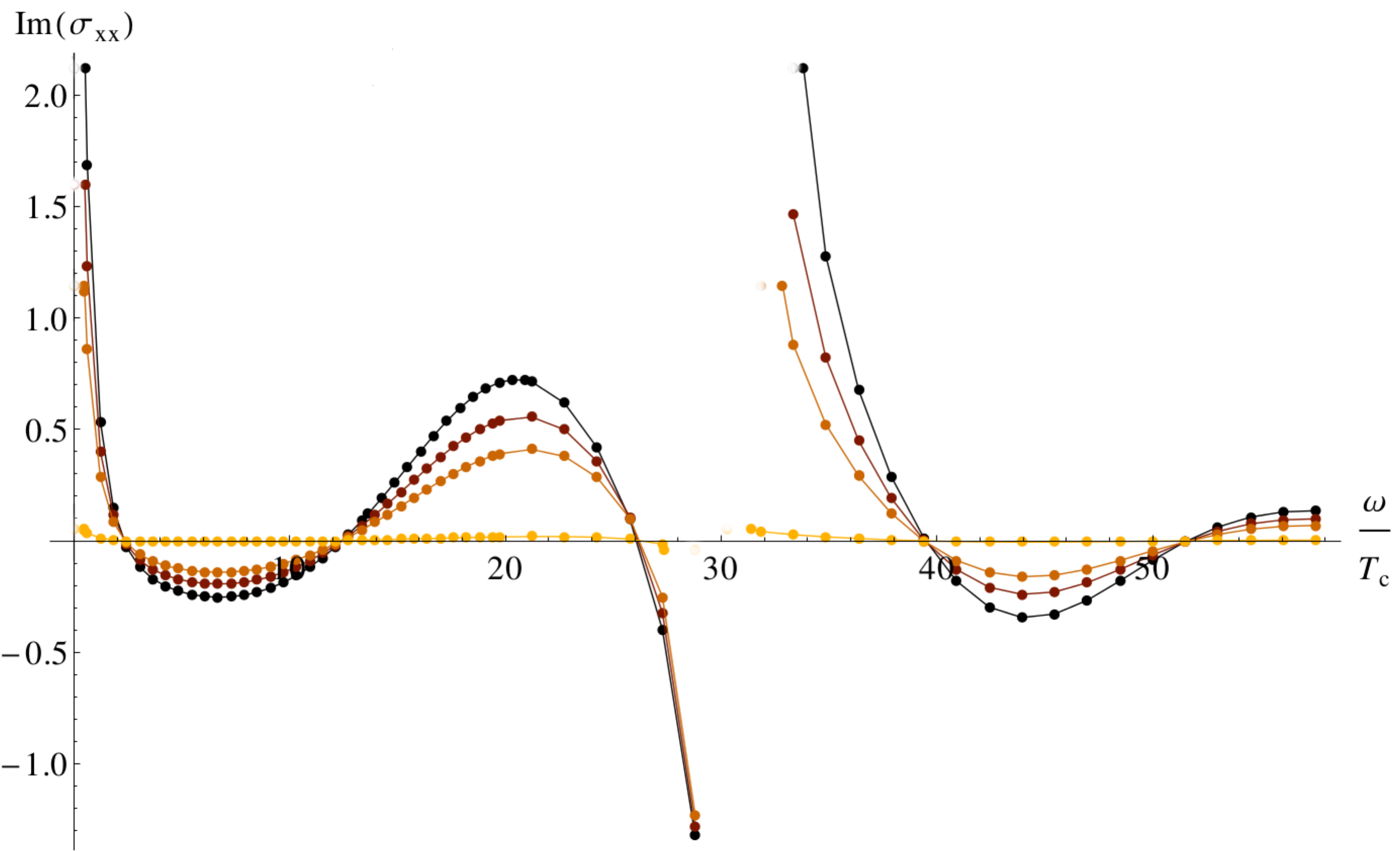} 
\caption{Real and imaginary part of the conductivity in the $xx$ direction at
decreasing temperature. The temperature is lowered from $T=0.99\ T_c$ (yellow) to $T=0.94\ T_c$, $T=0.92\ T_c$ 
and eventually $T=0.9\ T_c$ (black). This study concerning the transport has been performed over a background
characterized by $\mu_\phi = \mu_\eta = 1$. For the coupling constant we considered $c=1/10$. 
For different values of the constant $c$ we find that the behavior of the conductivity is qualitatively the same.}
\label{sigmacomplete}
\end{figure}
We define the conductivity in the $xx$ direction in the usual way, (see for example \cite{Hartnoll:2008vx}), as:
\begin{equation}
\label{sigma}
\sigma_{xx}=\frac{\hat{\bar{b}}_x^{3 \; (\text{VEV})} }{i \omega\ \hat{\bar{b}}_x^{3 (\text{source})}}\ .
\end{equation}
The numerical results for \eqref{sigma} are plotted in Figure \ref{sigmacomplete}. 
The first feature which should be noted is the presence of a pole both in the real and in the 
imaginary part of the conductivity.
The position of the pole falls in the interval $30 < \frac{\omega}{T} < 35$; its precise position 
depends on the temperature.
This pole is a robust feature of our model and it does not depend on the specific range of temperature%
\footnote{Indeed the pole is present also at temperature values which are very close to the critical point.};
furthermore, the pole never develops any imaginary part, then it does not signal an instability of our system. 
On a technical level, the presence of the pole appears to be connected with the denominator $\eta^2- \omega^2$ arising 
in the gauge invariant combination \eqref{gaugeinvb2}. We have successfully tested this statement checking that the position of 
the pole scales as the chemical potential $\mu_{\eta}$.

It is important to recall that an analogous pole was previously found also in the study of
the original p-wave holographic superconductor \cite{Gubser:2008wv}; 
there the numerical analysis showed a pole in the imaginary part of the $xx$ conductivity
(and then, by the Kramers-Kronig relations, a delta function in the real part) at a value of $\omega$ 
comparable to the value of the chemical potential. Furthermore, in a slightly different context,
a similar pole was found also in \cite{Iqbal:2010eh}. 
Both the above mentioned models share with the present system the characteristic of possessing a non-Abelian bulk gauge symmetry.
This is the crucial point. Indeed, the origin of the pole can be traced back to the presence of a denominator
which vanishes at $\omega = \mu$ and this denominator emerges in the construction of the 
gauge invariant bulk field combinations only for non-Abelian gauge structure. 
There is however a distinction between our results and those found in \cite{Gubser:2008wv}.
Actually, we find a pole both in the real and in the imaginary part of the conductivity. 
This discrepancy is probably due to the complicate gauge structure of our model and to the 
non-trivial interactions between the two vector fields $\bar{A}_{\mu}^a$ and $\bar{B}_{\mu}^a$.

The model presented in this paper is to be regarded in the spirit of ``holographic-effective'' theory. Said otherwise, it is expected 
to be valid only in the low-frequency regime where it describes the low-energy hydrodynamic limit of the
dual boundary theory. The phenomenological attitude, already implicit in adopting a bottom-up approach,
is more importantly motivated by the probe approximation. Indeed, our analysis relies on the possibility 
of neglecting the backreaction of the gauge fields on the metric. Such approximation cannot be valid for 
arbitrarily high values of the frequency (and then energy) of the fluctuating fields.

From a direct study of the equations of motion, we are not able to determine an upper
validity bound for $\omega$ as the probe equations do not encode the probe approximation on which 
they rely%
\footnote{Further analysis on a backreacted environment is already work in progress \cite{progress}.}.
Let us however observe that the position of the pole is $\omega/ T_c \sim 30$; this value indicates 
that we are probing the system with a frequency which is large when compared with its temperature.
Reading this fact the other way around, we have a temperature which is low with respect to the
characteristic scale of the fluctuations. And, again because of the probe approximation, our 
model cannot be considered reliable at low temperature.

\begin{figure}[ht]
\centering
\includegraphics[width=75mm]{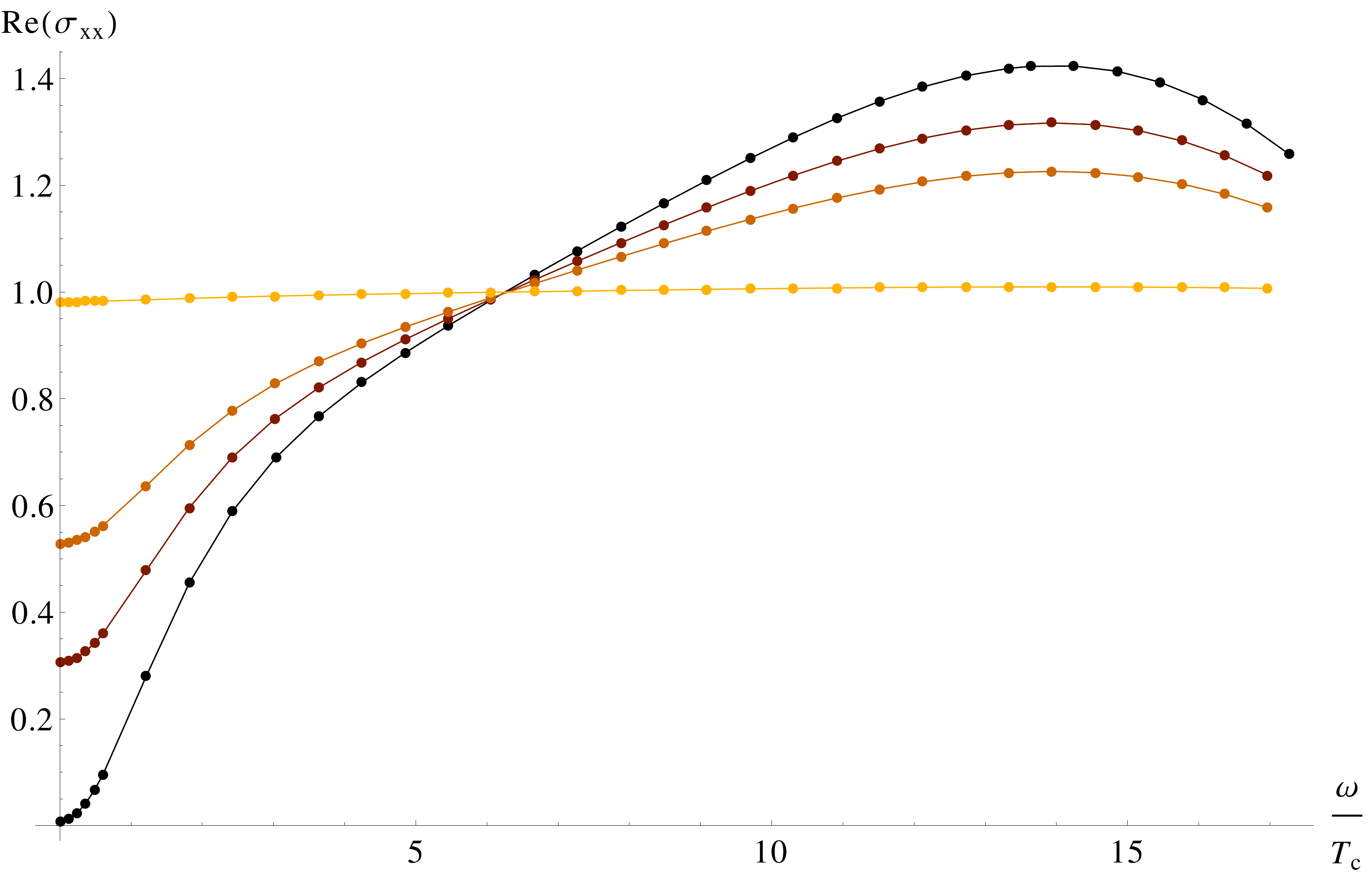} 
\includegraphics[width=75mm]{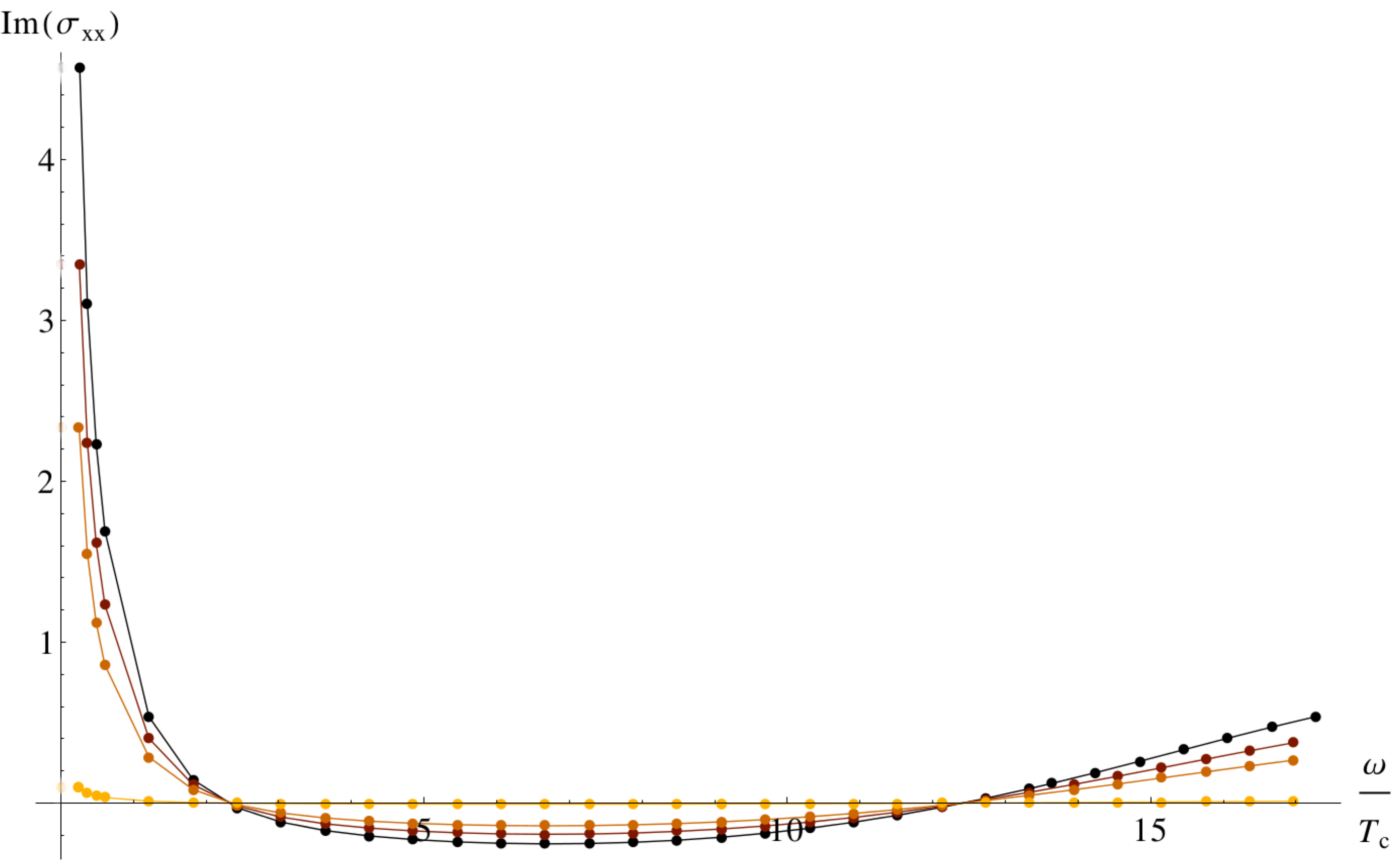} 
\caption{Low $\omega$ behavior of the real and imaginary part of the $xx$ conductivity for decreasing
values the temperature. Frequency is measured in terms of $T_c$. The yellow plot corresponds to $T=0.99\ T_c$, then we have $T=0.94\ T_c$, $T=0.92\ T_c$ 
and eventually the black line representing the case with $T=0.9\ T_c$. We remind the reader that the data
were obtained considering $c=1/10$ and $\mu_\phi / \mu_\eta=1$.}
\label{sigmaloww}
\end{figure}

We could trust our results in the low $\omega$ region defined as $\omega \sim \omega_{\text{gap}} \sim \mathcal{h} \mathcal{O}_v \mathcal{i} \ll \mu_{\eta}$.
The optical conductivity in such a range is illustrated in Figure \ref{sigmaloww}.
The imaginary part of the conductivity shows a pole in $\omega=0$; consequently, the real part presents there
a delta function. 
As expected, the pole amplitude decreases with the increasing of the temperature since the coefficient of the pole in the imaginary
part of the conductivity is the superfluid density $n_s$, ($Im[\sigma_{xx}] \sim \frac{n_s}{\omega}$ for $\omega \rightarrow 0$).
In the low-frequency region, a gap appears in the real part of the conductivity when the temperature is lowered. 
In Figure \ref{sigmaloww}, the line featuring the most pronounced gap is obtained for $T=0.9\, T_c$, where
the real part, $Re[\sigma_{xx}]$, is very small in the deep infrared and then grows quickly around $\omega \sim 2 \mathcal{h} \tilde{\mathcal{O}}_v \mathcal{i}$.

Finally, we note that there is a change in the slope of the real part of the conductivity for $2 < \frac{\omega}{T_c} <5$, 
when $\omega$ is comparable with the spin density order parameter $\mathcal{h} \tilde{\mathcal{O}}_w \mathcal{i}$. This furnishes evidence of the fact that 
the system has two relevant scales related to the amplitude of the two condensates $\mathcal{h} \tilde{\mathcal{O}}_v \mathcal{i}$ 
and $\mathcal{h} \tilde{\mathcal{O}}_w \mathcal{i}$ and 
that the two vector fields $\bar{A}_{\mu}^a$ and $\bar{B}_{\mu}^a$, which we have interpreted as a spin density order 
parameter and a superconducting order parameter respectively, interact with each other.

In conclusion, we have shown that, in the low-$\omega$ hydrodynamic limit, the model reproduces the behavior expected from a
superconductor and that there is evidence for the presence of two relevant scales in the system, (the superconducting and 
the spin one). This will become more evident in the next Section where we shall study the susceptibility of the model.

Forgetting the pole structure discussed before, the high-frequency (i.e. $\omega\gtrsim 35$) behavior of the conductivity $\sigma_{xx}$ corresponds to
a constant unitary value as expected on general grounds in holographic systems \cite{Herzog:2007ij}. 

\paragraph{The transverse sector} $\;$\\
As we have illustrated at the beginning of Section \ref{traconde}, in the transverse sector the spin and electric responses
are mixed and described with the conductivity matrix \eqref{transportmatrix}. 
Then, in line with the proposed phenomenological interpretation, we read the diagonal entry 
$\sigma_{\bar{b} \bar{b}}(\omega)$ in \eqref{transportmatrix} as the pure electric transverse conductivity $\sigma_{yy}(\omega)$.
\begin{figure}[ht]
\centering
\includegraphics[width=75mm]{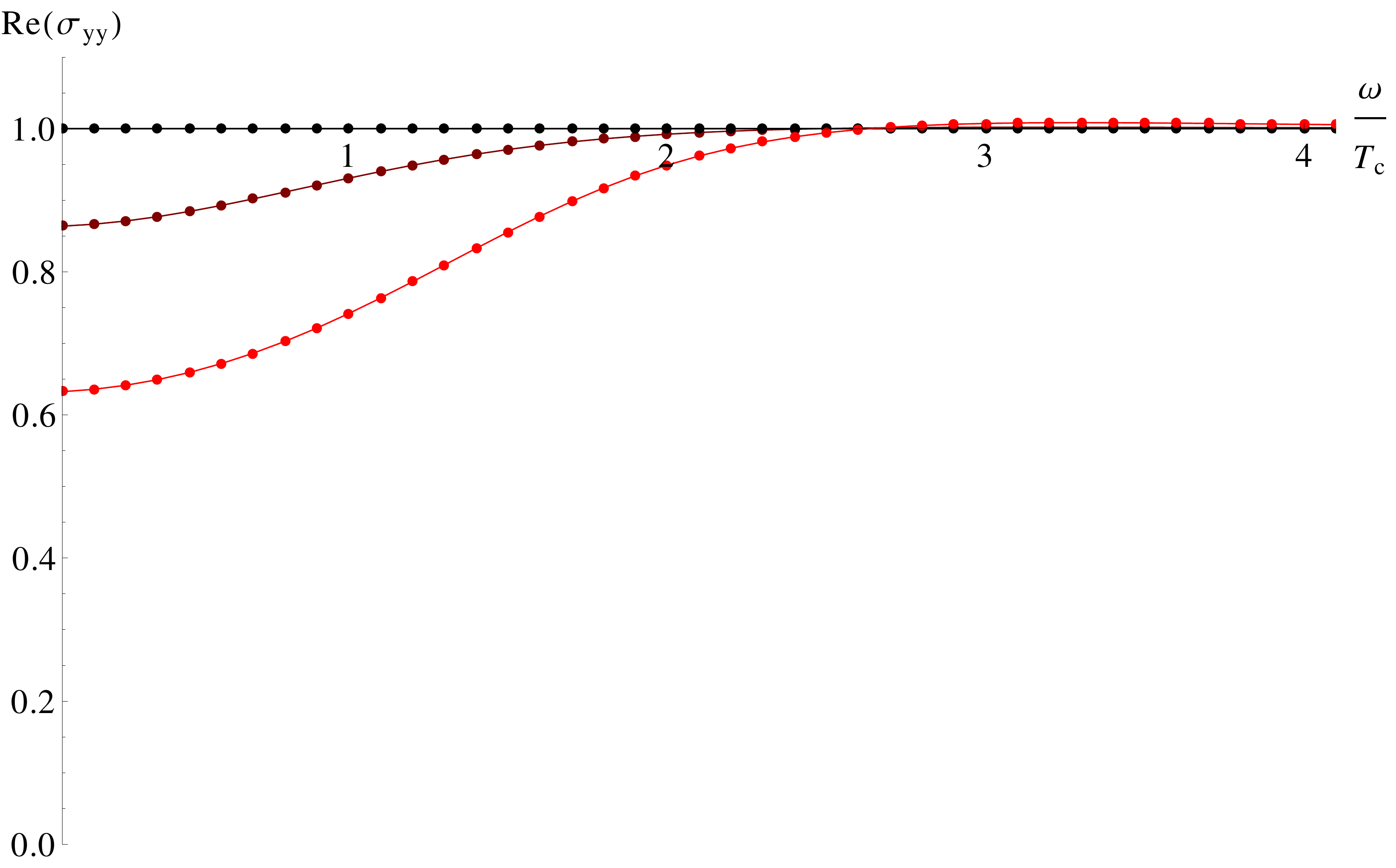} 
\includegraphics[width=75mm]{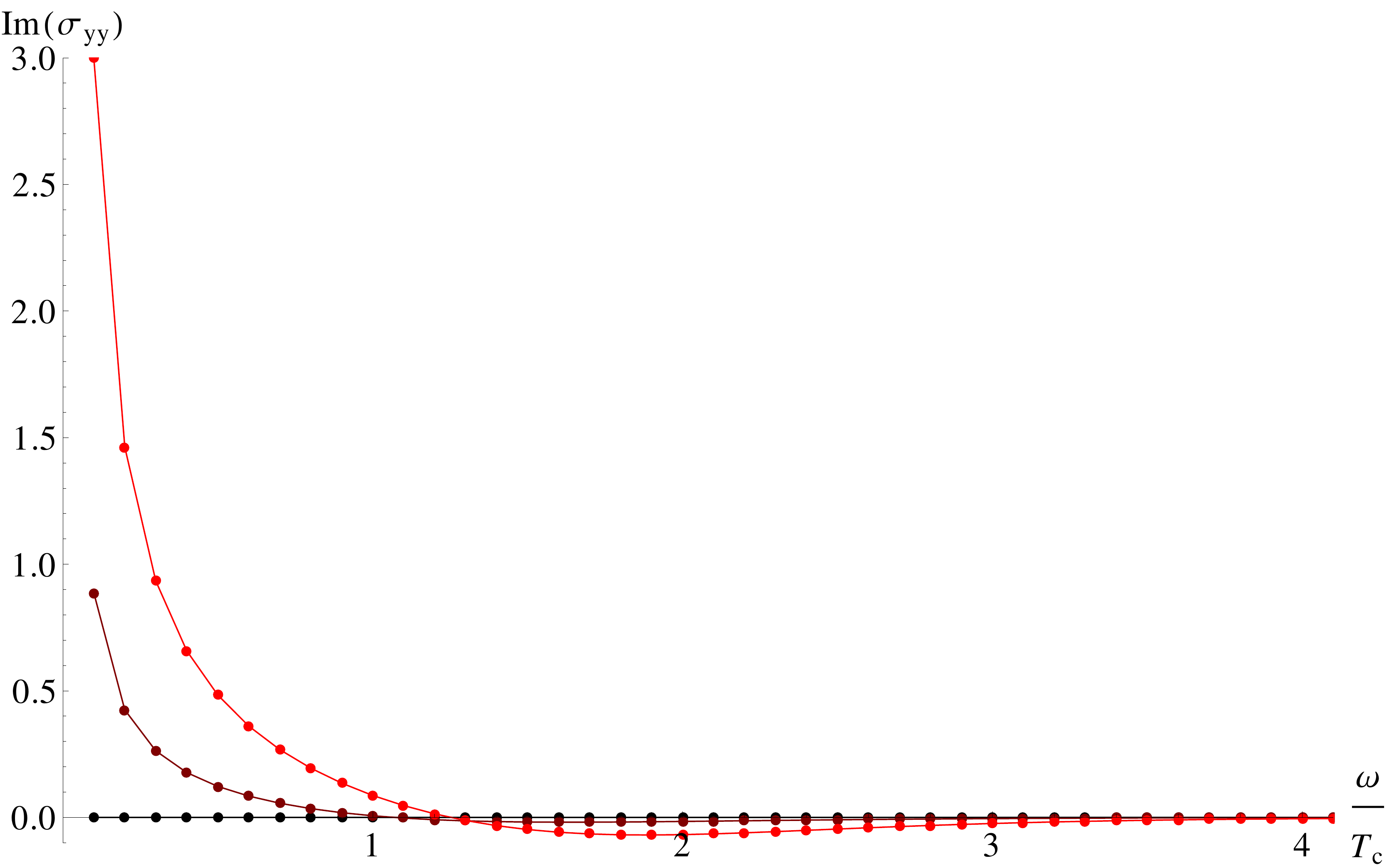} 
\caption{Low $\omega$ behavior of the real and imaginary part of the $yy$ conductivity for $T=0.99\ T_c$ (black), $T=0.97\ T_c$ (brown) and $T=0.90\ T_c$ (red).  The data
were obtained considering $c=1/10$ and $\mu_\phi/\mu_\eta=1$. For different values of the constant $c$ the behavior of $\sigma_{yy}$ is qualitatively the same.}
\label{sigmalowwtrasv}
\end{figure}
The numerical results for $\sigma_{yy}$ at decreasing temperature are reported in Figure \ref{sigmalowwtrasv}. 
The most important feature to note is that, for $T=0.9\, T_c$, $\sigma_{yy}$ is significantly different from 
zero while the gap in $\sigma_{xx}$ is noticeable and the corresponding conductivity is approximately null. 
This is similar to the behavior of the $\sigma_{yy}$ found in the holographic p-wave 
superconductor studied in \cite{Gubser:2008px}, and is a feature which is possible to find in an ordinary 
p-wave superconductor due to the anisotropy of the gap, suggesting that the gap is in the $xx$ direction \cite{basov2005,lee2006}. 

\subsubsection{The susceptibility} 

Inspired by the ferromagnetic superconductor interpretation, it is interesting to compute
the spin susceptibility of the present holographic model in order to characterize 
more precisely the features exhibited by its ferromagnetic ordering.

At first, we review some basic notions about the spin susceptibility in linear response theory%
\footnote{For more details about this subject see, for example, \cite{Kadanoff1963419}.}. 
If we consider a system immersed in a weak external field $H^i(x_{\mu})$ which couples directly with the spin density, its Hamiltonian 
$\mathcal{H}$ is perturbed by the term
\begin{equation}
\delta \mathcal{H}=\int d^dx\ S_i(x_{\mu}) H^i(x_{\mu}),
\end{equation}
where $S_i(x_{\mu})$ is the spin density along the $i$-th spatial direction. 
At linear order, the external field induces the following variation of the spin density
expectation value
\begin{equation}
\label{linearresponse}
\delta \mathcal{h} S_i(x_{\mu}) \mathcal{i}=\int d^dx\ G_{ij}^R (x_{\mu},x'_{\mu}) \delta H^j(x'_{\mu}),
\end{equation}
where $G_{ij}^R (x_{\mu},x'_{\mu})$ is the retarded Green's function defined as follows
\begin{equation}
G_{ij}^R (x_{\mu},x'_{\mu})=-i \theta(t-t') \mathcal{h} \left[S_i(x_{\mu}),S_j(x'_{\mu}) \right] \mathcal{i}.
\end{equation}
Taking the Fourier transform of equation \eqref{linearresponse} it is easy to see that the spin susceptibility $\chi_{ij}=\frac{\delta S_i}{\delta H^j}$ 
is directly given by the retarded Green function
\begin{equation}
\label{chichi}
\chi_{ij}(\omega,\vec{k})=\tilde{G}^R_{ij}(\omega,\vec{k})=\int d^dx\ e^{-i \omega t+i k^l x_l}\ G_{ij}^R(x_{\mu},0)\ .
\end{equation}
The static susceptibility is defined as the limit:
\begin{equation}
\chi_{ij}^{(\text{static})}=\lim _{\omega \rightarrow 0 }  \left[ \lim_{ \vec{k} \rightarrow 0} \tilde{G}^R_{ij}(\omega,\vec{k}) \right],
\end{equation}
where the long wavelength limit is taken before the limit $\omega \rightarrow 0$.
Returning to the holographic model at hand, in the previous sections we have shown how 
the gauge structure of the model allows us to interpret 
the order parameter associated to the field $\bar{A}_{\mu}^a$ in terms of a spin density order parameter. As a consequence, the fluctuations of the bulk field $\bar{A}_{\mu}^a$ 
(indicated with $\bar{a}^a_{\mu}$) are associated to the spin fluctuations around the 
boundary theory equilibrium. In agreement with the previous analysis of the bulk fluctuations,
we interpret their gauge invariant combinations in connection with the observables 
of the boundary field theory. 

\paragraph{Numerical results} $\;$\\
We focus our attention on the spin susceptibility directed along the same direction of the 
ferromagnetic order parameter $A_x^3$, namely, the $x$ direction. As it emerged in Section 
\ref{gaugeinvariant}, the only gauge invariant combination involving $\bar{a}^a_{\mu}$
which is along the $x$ direction is 
$\hat{\bar{a}}_x^3=\bar{a}_x^3+\frac{w}{\phi}\bar{a}_t^1$, (see \eqref{gaugeinvb1}).
Consequently, it is natural to interpret the correlator 
$\mathcal{h}\hat{\bar{a}}_x^3(x')\hat{\bar{a}}_x^3(x) \mathcal{i}$
as the spin-spin correlation function:
\begin{equation}
\mathcal{h}\hat{\bar{a}}_x^3(x')\hat{\bar{a}}_x^3(x) \mathcal{i}=-i\theta(t-t') \mathcal{h}\left[ S_x(x'),S_x(x) \right] \mathcal{i}=G^R_{xx}(x,x').
\end{equation}
Eventually, keeping in mind the definition \eqref{chichi}, the spin susceptibility 
of the present holographic model may be defined as
\begin{equation}\label{susce}
\chi_{xx}(\omega,\kappa) \equiv \mathcal{h}\hat{\bar{a}}_x^3(\omega, \kappa)\hat{\bar{a}}_x^3(-\omega, -\kappa) \mathcal{i},
\end{equation}
where $\kappa \equiv k_x$%
\footnote{We remind the reader that, in the discussions contained in the previous sections,
we have set $k_y=0$ and $\vec{k}=k_x \equiv \kappa$.}. 
From the definition of the spin susceptibility \eqref{susce},
it is straightforward to compute it in accordance with the standard holographic prescription
(see, for example, \cite{Son:2002sd}), namely
\begin{equation}\label{chichichi}
\chi_{xx}(\omega, \kappa)=\frac{\hat{\bar{a}}_x^{3\;(\text{VEV})}}{\hat{\bar{a}}_x^{3\;(\text{source})}}\ .
\end{equation}
\begin{figure}[ht]
\centering
\includegraphics[width=75mm]{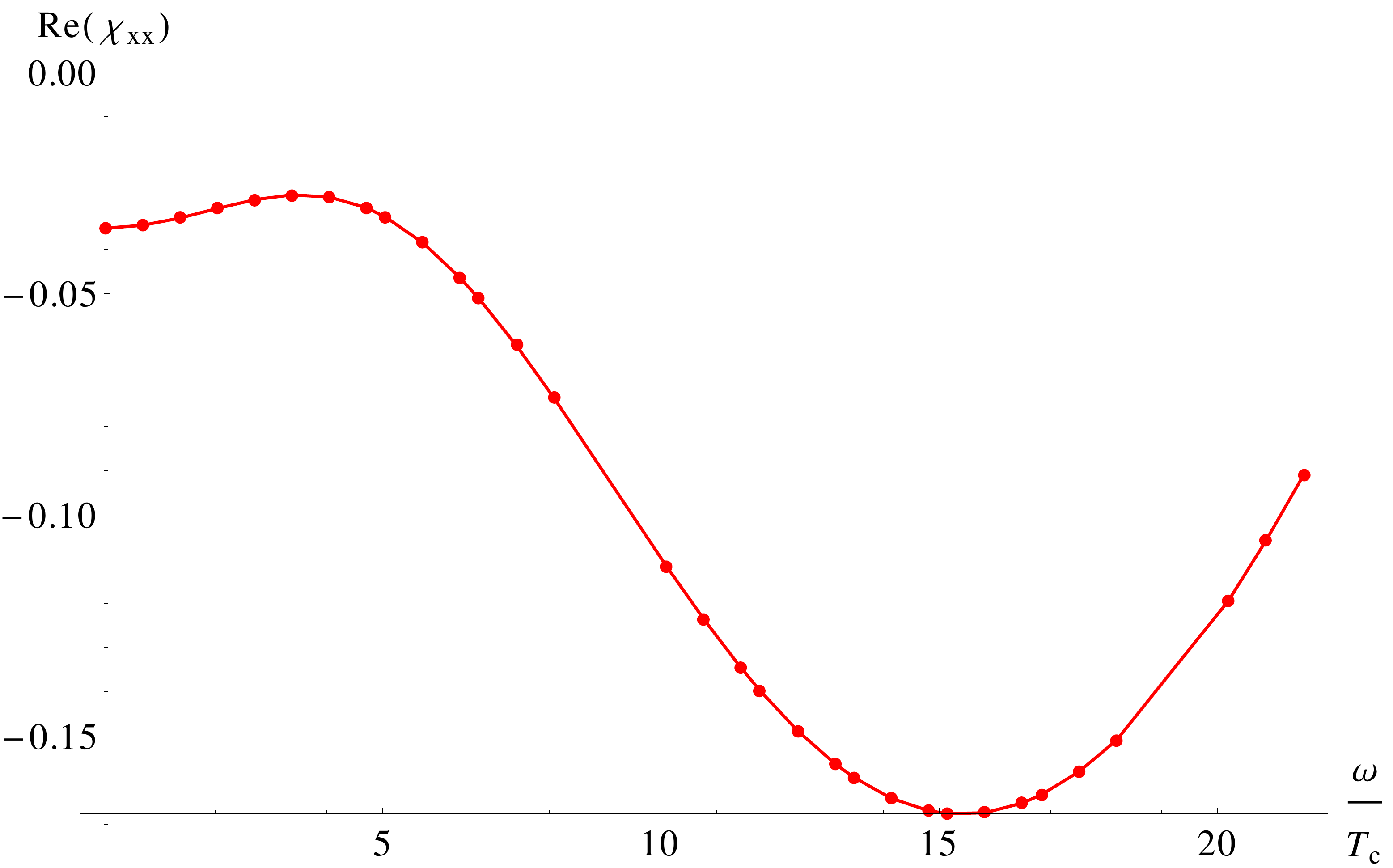} 
\includegraphics[width=75mm]{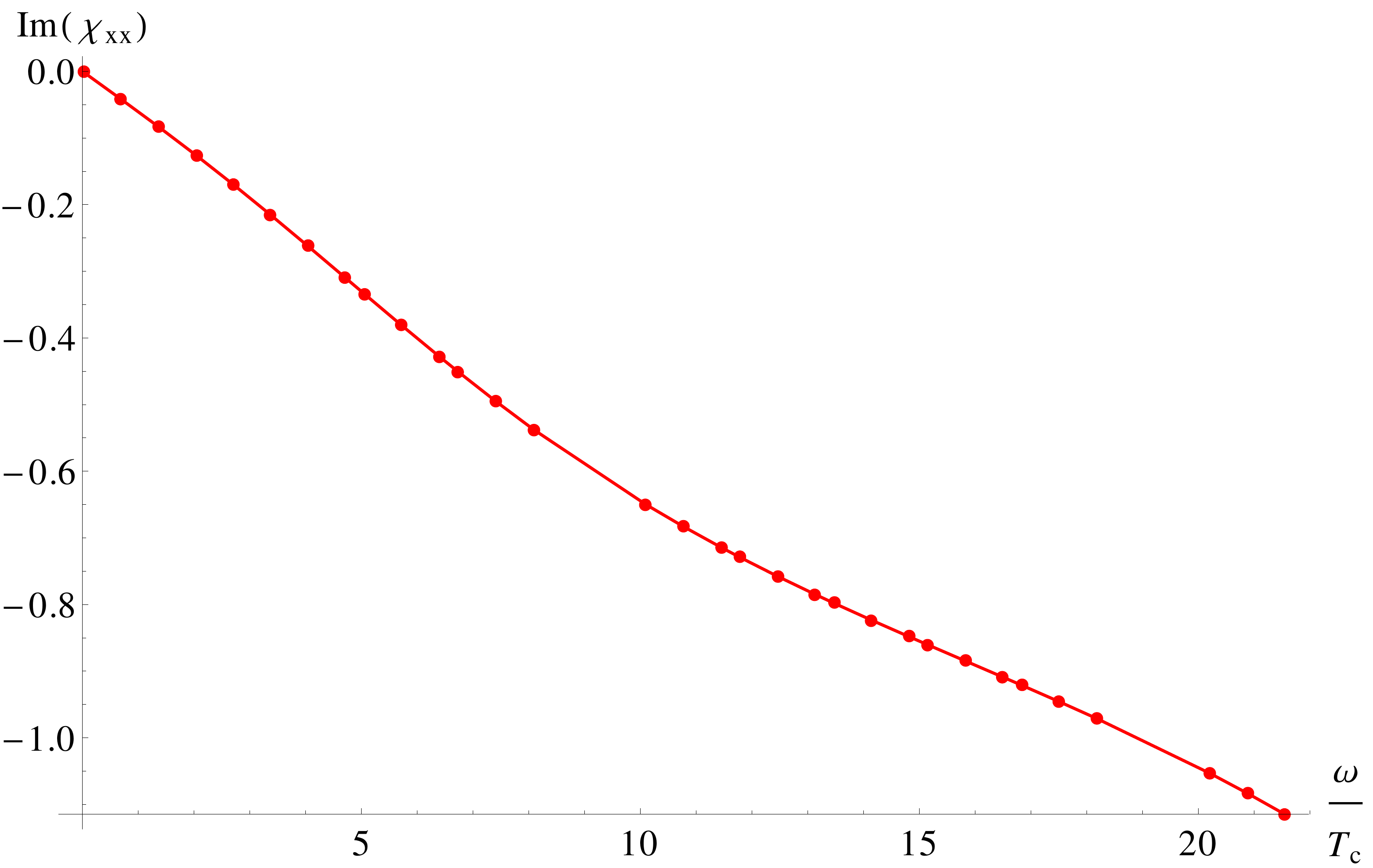} 
\caption{The real and the imaginary part of the longitudinal susceptibility (defined in \eqref{chichichi})
as a function of $\frac{\omega}{T_c}$ for $T=0.9T_c$.}
\label{chiw}
\end{figure}

Let us first analyze the AC susceptibility $\chi_{xx}(\omega,\kappa=0)$, whose real and 
imaginary parts are plotted in Figure \ref{chiw} for $T=0.9\, T_c$. 
As it is evident from the left plot in Figure \ref{chiw}, the low $\omega$ regime 
presents two changes in the slope of the real part of the AC susceptibility: 
the first one corresponds to $\frac{\omega}{T_c} \sim 1$ and the second one is around $3 <\frac{\omega}{T_c}<5$. 
Remarkably, such changes in the slope occur in the same interval of $\frac{\omega}{T_c}$ in which the AC conductivity, 
(evaluated at the same temperature), changes its slope as well, as illustrated in the previous section. 
As anticipated in the analysis of the conductivity, this provide further evidence of the fact that
there are two relevant scales in the system, (namely the two order parameters $ \mathcal{h} \tilde{\mathcal{O}}_v \mathcal{i}$ 
and $\mathcal{h} \tilde{\mathcal{O}}_w \mathcal{i}$), 
and that the spin and the charge sectors of our model interact with each other. In other terms,
both the electric and spin responses are sensitive to both scales. 
This qualitative behavior does not change for different values of the coupling constant $c$.

We further observe that the imaginary part of the susceptibility (right plot in Figure \ref{chiw})
has no pole at $\omega=0$ and therefore, according with the Kramers-Kronig relations, the real part 
of the static susceptibility is finite (i.e. there is no DC delta function in the real part 
of the susceptibility). Phenomenologically, this agrees with the fact that, in a ferromagnet, 
\begin{figure}[ht]
\centering
\includegraphics[width=75mm]{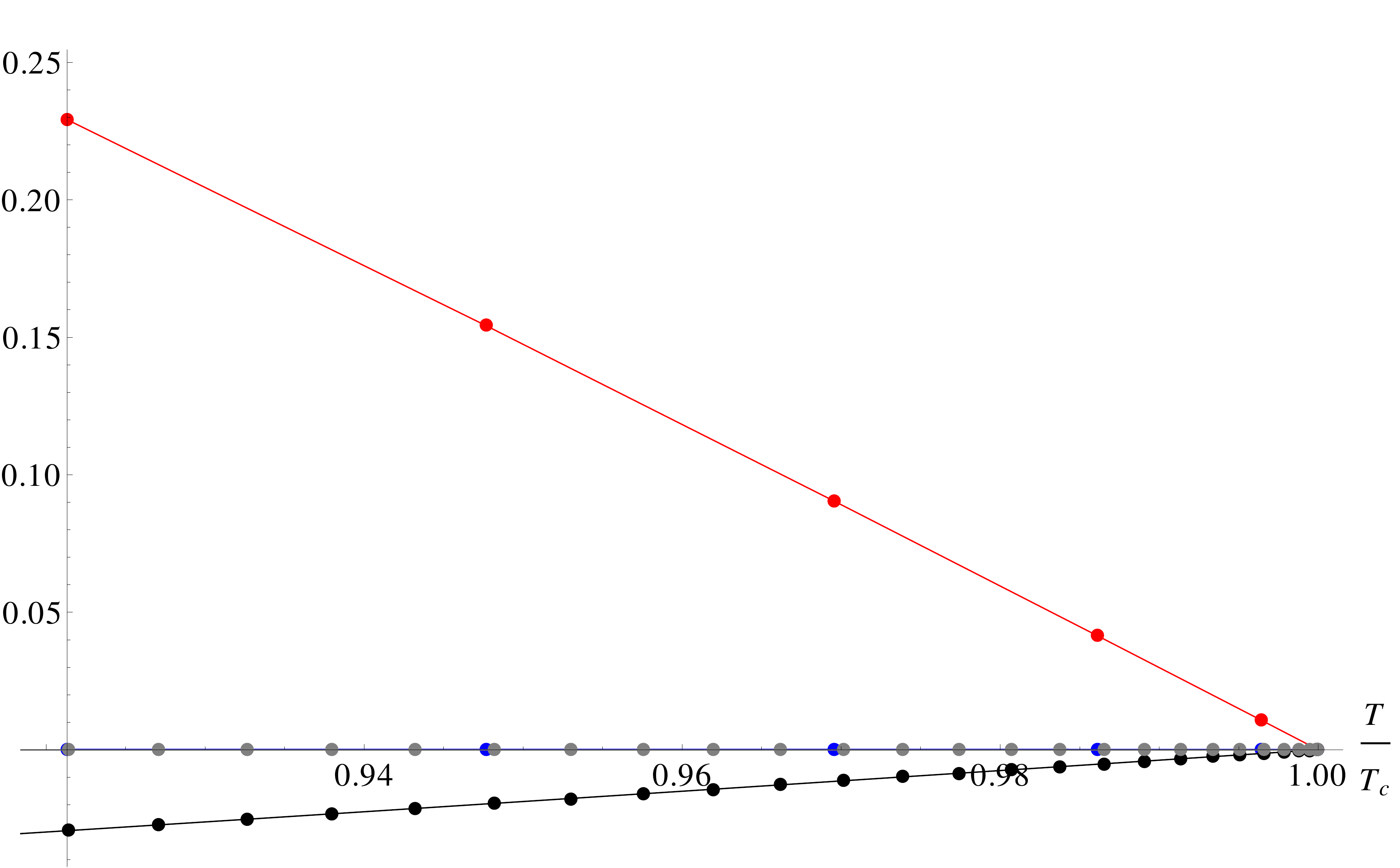} 
\includegraphics[width=75mm]{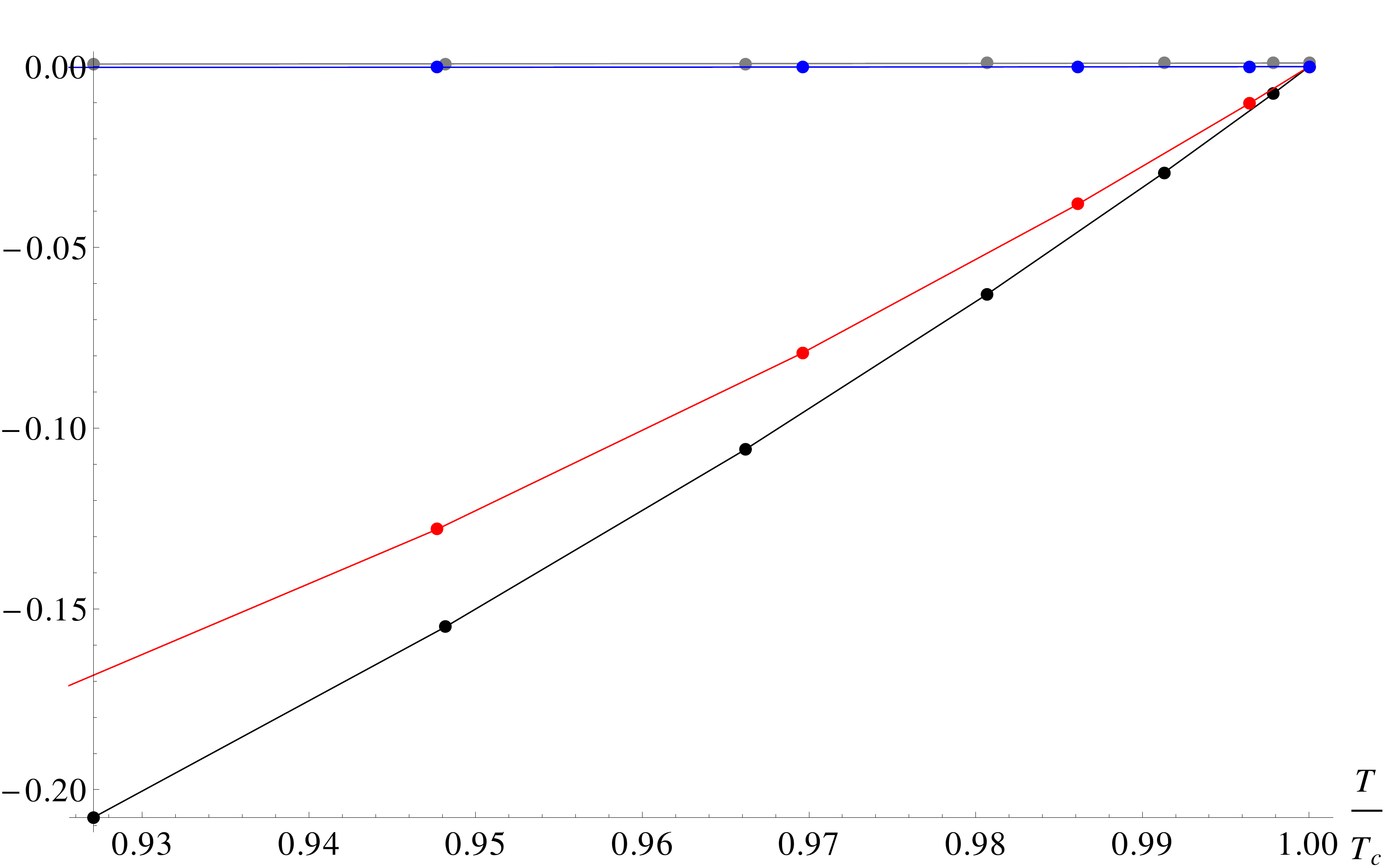} 
\caption{Real and imaginary parts of the static susceptibility vs the normalized temperature for the longitudinal sector (left) and the transverse sector (right). 
For $c=\frac{1}{10}$ the real part is represented by the black line and the imaginary part by the gray line. 
For  $c=-\frac{3}{10}$ the real part is represented by the red line and imaginary part by  the blue line.}
\label{chistati}
\end{figure}
the spin susceptibility along the magnetization direction must be positive and finite at finite temperature. 
Furthermore, we note that the imaginary part of the static susceptibility is zero, which, remarkably, 
is the same value predicted by the generalized BCS theory \cite{sigrist}.

It is interesting to study the behavior of the static susceptibility as a function of the 
temperature; we report in Figure \ref{chistati} the results for $c=\frac{1}{10}$ and $c=-\frac{3}{10}$. 
Remarkably, depending on the sign of $c$, $Re \left(\chi_{xx} \right)$ starts from zero at $T_c$ and linearly 
decreases for $c>0$ or linearly increases for $c<0$. 
This seems reasonably connected to the fact that, when the ferromagnetic order parameter dominates the 
superconducting one (as occurs for $c<0$, see Figure \ref{condensates}), the longitudinal susceptibility is 
dominated by the ferromagnetic behavior. In the opposite case, when $c>0$, the superconducting order parameter 
is dominating and induces a diamagnetic behavior.

An intrinsic characteristic of the ferromagnetism modeled by the present holographic system is 
the feature of being itinerant. The ferromagnetic 
order that we observe is not related to any fixed or static degree of freedom, indeed, our
holographic model has no spatial features that break translation invariance (as, for instance, a
lattice). This itinerant nature is exactly what is experimentally suggested to hold for 
UCoGe, URhGe and $\text{UGe}_2$ \cite{Aoki:2001}.

More precisely, observe that the present model has some analogy 
with the minimal unbalanced holographic superconductor introduced in \cite{Bigazzi:2011ak}.
In particular, in the normal phase, they are described by the same $AdS$-RN solution. 
In \cite{Bigazzi:2011ak} it has been shown that the normal phase conductivity matrix can be parametrized
in terms of a single $\omega$-dependent mobility function. Intuitively, this means that the linear response 
of the system has an analogous structure as if the transport were due to ``carriers'' charged both electrically 
and magnetically. In other terms, the spin and electric properties of these holographic systems
appear to be related to the same itinerant degrees of freedom.

If we consider the transverse susceptibility $\chi_{yy}$, following the argument of the previous paragraphs, 
we interpret the diagonal entry $\chi_{\bar{a} \bar{a}}(\omega)$ of the transport matrix \eqref{transportmatrix} as the transverse susceptibility $\chi_{yy}$. 
The numerical data for $\chi_{yy}^{\text{(static)}}$ are plotted in Figure \ref{chistati} for $c=\frac{1}{10}$ and $c=-\frac{3}{10}$. 
It is interesting to note that, unlike $\text{Re} \left(\chi_{xx}^{\text{(static)}} \right)$, 
$\text{Re} \left(\chi_{yy}^{\text{(static)}} \right)$ decreases with the temperature both for 
$c>0$ and for $c<0$, showing a generic diamagnetic behavior in the transverse channel, independently of the sign of $c$. 

\section{Conclusion and future prospects}

We have introduced and investigated a holographic model 
featuring the coexistence of two vector order parameters. The model
represents a natural extension of the standard holographic p-wave 
superconductor \cite{Gubser:2008wv} possessing two non-Abelian vector fields
whose mutual interactions are determined by a bilinear term in their 
field strengths. Such an interaction term forces (by gauge invariance) to have
identical gauge transformations for both field strengths. Once the kinetic terms
are diagonalized by means of a rotation, the rotated fields are related essentially
to the sum and the difference of the original vector fields. This property is crucial
in determining the gauge transformation properties of the two rotated fields.
It turns out that we have an authentic gauge field plus vectorial matter 
in the adjoint representation.

In this paper we have considered a non-Abelian $SU(2)$ gauge invariance.
This is explicitly broken by a restrictive ansatz which leads to a $U(1)\times U(1)$
``residual'' symmetry emerging from an explicit breaking $SU(2)\rightarrow U(1)$
combined by a doubling effect arising from the Abelianization of our model
under the ansatz. The two Abelian factors allow for an interpretation analogous to
the treatment of the unbalanced holographic superconductor \cite{Bigazzi:2011ak}
where one $U(1)$ is interpreted as the charged degrees of freedom whose breaking leads to 
superconductivity while the second $U(1)$ accounts effectively for spin degrees
of freedom. The spontaneous breaking of the latter is then interpreted as a 
ferromagnetic phenomenon.
The present system features a concomitant condensation of the two order parameters,
indicating that the ferromagnetism and the p-wave superconductivity occur at the same energy scale.

The double condensation is a second order transition, as the study of the free energy at the critical point confirms. The phase diagram is rich and its 
qualitative structure is independent of the strength of the coupling $c$.%
\footnote{At least as long as $c\neq0$. Actually, the $c=0$ case is particular as the 
two original vectors of the model do not have direct interactions. In a probe analysis as
the one presented here they would have no interaction at all. Note that this trivializes 
the model reducing it to just a double copy of the standard p-wave holographic superconductor.}
In particular it features a first order transition between two doubly-condensed phases distinguished
by a relative sign among the condensates. 
The coupling $c$ assumes the crucial role of a free parameter that determines the universality class of the properties reported in this paper.

The charge transport and spin susceptibility  were analyzed within the paradigm 
of linear response theory. The superconductivity gap
shows a structure which may be related to the triplet p-wave anisotropic nature of the superconducting mechanism. 
Beside the gap, the optical conductivity shows the signature of the presence of the spin density order parameter. 
The longitudinal susceptibility shows a direct connection with the strength of the order parameters, 
being positive when the ferromagnetic order dominates 
($c<0$, $\mathcal{h}\mathcal{O}_w\mathcal{i}>\mathcal{h}\mathcal{O}_v\mathcal{i}$), 
and weakly diamagnetic when the superconductivity order takes over 
($c>0$, $\mathcal{h}\mathcal{O}_w\mathcal{i}<\mathcal{h}\mathcal{O}_v\mathcal{i}$).

The conductivity shows a pole in the $\omega/T_c$ region above the gap that is 
associated to a non-dissipative, massive mode arising from the non-Abelian gauge structure 
of the model. However, in an effective theory spirit, this structure is observed in a frequency 
region where the probe approximation is not likely to be appropriate anymore. 
The study of this pole, including the backreaction will be an interesting direction for the future work%
\footnote{The analysis of the equilibrium of the standard holographic p-wave superconductor 
taking into account the backreaction is performed in \cite{Ammon:2009xh,Arias:2012py}.}.
A backreacted environment is necessary to pose soundly the interpretation of the higher frequency region of the optical and eventually thermal transport.
One direction for future investigation concerns also the stability of the condensed phase in a way similar 
to what was done for the standard p-wave superconductor (see \cite{Gubser:2008px} for further details).

We hope that this analysis will trigger the experimental community to investigate better the regime 
where the two order parameters have almost the same energy scale, analyzing the intriguing regime of
interplay and feedback between triplet superconductivity and magnetism.

\section{Acknowledgements}
DM and AA want to thank Andrea Mezzalira, Irene Amado, Christopher Herzog,
Matthias Kaminski,  Daniel Arean and Javier Tarrio for very useful discussions.
DM would like to thank also Aldo Cotrone, Francesco Bigazzi, Alberto Lerda, Davide Forcella, 
Ignacio Salazar Landea, Diego Redigolo, Manuela Kulaxizi, Rakibur Rahman, Micha Moskovic, 
Gustavo Lucena Gomez, Sean Hartnoll, Joe Bhaseen, Katherine Michele Deck and Giorgio Musso 
for their important suggestions and interesting discussions.

The work of DM is partially supported by IISN - Belgium (conventions 
4.4511.06 and 4.4514.08), by the ``Communaut\'e Fran\c{c}aise de 
Belgique" through the ARC program and by the ERC through the ``SyDuGraM" 

We also thank Mario Cuoco, Carlo Maria Becchi, the support of INFN Scientific Initiative: 
"Statistical Field Theory, Low-Dimensional Systems, 
Integrable Models and Applications" and the FIRB2012 - Project HybridNanoDev RBFR1236VV
and EU FP7 Programme under Grant Agreement No. 234970-NANOCTM.

A.B. acknowledge the hospitality of the Institute for Nuclear 
Theory (INT-PUB-13-040) in Seattle where the work was partially done.

\appendix

\section{Linearized equations and constraints for the fluctuations}
\label{lineeq}
The twelve equations of motion for the longitudinal fluctuations expressed in the unphysical (i.e. unrotated) basis are:
\begin{multline}
a_t^{1''}-c\, b_t^{1''} +\frac{2 \left(a_t^{1'}-c\, b_t^{1'}\right)}{r}-\frac{\kappa ^2 \left(a_t^1-c\, b_t^1\right)}{h\, r^2}\\
-\frac{i \kappa 
   \left(c H\,  b_x^2-\Phi\,  a_x^2\right)+\kappa\,  \omega 
   \left(a_x^1-c\, b_x^1\right)+a_x^3 (c H  V-W \Phi
   )}{h\, r^2}=0,
\end{multline}
\begin{multline}
a_t^{2''}-c\, b_t^{2''}+\frac{2 \left(a_t^{2'}-c\, b_t^{2'}\right)}{r}-\frac{\kappa ^2 \left(a_t^2-c\, b_t^2\right)}{h\, r^2}-\frac{i \kappa 
   \left[2 W a_t^3-c\, (V+W) b_t^3\right]}{h\, r^2}\\
  -\frac{i W \omega    \left(a_x^3-c\, b_x^3\right)}{h\, r^2} -\frac{i \kappa  \left(\Phi \,
   a_x^1-c\, H  b_x^1\right)+\kappa \, \omega 
   \left(a_x^2-c\, b_x^2\right)+W^2 a_t^2-c\, V W
   b_t^2}{h\, r^2}=0,
\end{multline}
\begin{multline}
a_t^{3''}-c\, b_t^{3''}+\frac{2 \left(a_t^{3'}-c\,
   b_t^{3'}\right)}{r}-\frac{\kappa ^2 \left(a_t^3-c\, b_t^3\right)+W^2
   a_t^3+2 W \Phi  a_x^1-c V W b_t^3-c H  W
   b_x^1}{h\, r^2}\\
-\frac{i \kappa 
   \left[c\, (V+W) b_t^2-2 W a_t^2\right]+i W \omega 
   \left(c\, b_x^2-a_x^2\right)+\kappa  \omega 
   \left(a_x^3-c b_x^3\right)-c H  V a_x{}^1}{h\, r^2}=0,
\end{multline}
\begin{multline}
a_x^{1''}-c\, b_x^{1''}+\frac{h'\left(a_x^{1'}-c\,  b_x^{1'}\right)}{h}+\frac{c H  \left(-V
   a_t^3+i \omega  b_x^2\right)+\omega ^2 \left(a_x^1-c\,
   b_x^1\right)+\kappa\,  \omega  \left(a_t^1-c\, b_t^1\right)}{h^2}\\
+\frac{\Phi 
   \left(-i \kappa\, a_t^2+2 W a_t^3-2 i\, \omega \,
   a_x^2+\Phi  a_x^1+i c\, \kappa\,  b_t^2-c\, V b_t^3-c\,
   H  b_x^1+i c\, \omega  b_x^2\right)}{h^2}=0, 
\end{multline}
\begin{multline}
a_x^{2''}-c b_x^{2''}+\frac{h' \left(a_x^{2'}-c\,
   b_x^{2'}\right)}{h}+\frac{\kappa \, \omega  \left(a_t^2-c\, b_t^2\right)+\omega ^2 \left(a_x^2-c\, b_x^2\right)}{h^2}\\
+\frac{\Phi \left(i \kappa \, a_t^1+2 i\, \omega\,  a_x^1+\Phi 
   a_x^2-i c\, \kappa\,  b_t^1-c H  b_x^2-i c\, \omega \,
   b_x^1\right)}{h^2}\\
   +\frac{i
   W \omega\,  a_t^3-i\, c\, V\, \omega\,  b_t^3-i c H  \omega\, 
   b_x^1}{h^2}=0,
\end{multline}
\begin{multline}
a_x^{3''}-c\, b_x^{3''}+\frac{h' \left(a_x^{3'}-c\,
   b_x^{3'}\right)}{h}+\frac{\kappa  \omega  \left(a_t^3-c\, b_t^3\right)+\omega ^2
   \left(a_x^3-c\, b_x^3\right)}{h^2}\\
+\frac{i \omega    \left(c\, V b_t^2-W a_t^2\right)+c\, H  V a_t^1-W \Phi 
   a_t^1}{h^2}=0,
\end{multline}
and the other six are obtained by the previous exchanging $a_{\mu}^a \leftrightarrow b_{\mu}^a$, 
$W \leftrightarrow V$ and $H \leftrightarrow \Phi$%
\footnote{Recall the exchange of the original unrotated fields $A \leftrightarrow B$ is a symmetry of the model.}.

The six constraints are:
\begin{multline}
-c\, r^2 H ' a_t^2-i\, h\, \kappa\,  a_x^{1'}-i r^2
   \omega\,  a_t^{1'}+r^2 \Phi ' a_t^2-r^2 \Phi 
   a_t^{2'}+i c\, h\, \kappa 
   b_x^{1'}+
   i c\, r^2 \omega 
   b_t^{1'}+c\, r^2 \Phi  b_t^{2'}=0,
\end{multline}
\begin{multline}
c\, h\, V' a_x^3+c r^2 H ' a_t^1-h W' a_x^3+h W
   a_x^{3'}-i h \kappa  a_x^{2'}-i
   r^2 \omega  a_t^{2'}\\
   -r^2 \Phi ' a_t^1+r^2
   \Phi  a_t^{1'}-c\, h\, W\, b_x^{3'}+i
   c\, h\, \kappa\,  b_x^{2'}+i c\, r^2 \omega 
   b_t^{2'}-c\, r^2 \Phi  b_t^{1'}=0,
\end{multline}
\begin{multline}
-c h V' a_x^2+h W' a_x^2-h W a_x^{2'}-i h
   \kappa  a_x^{3'}-i r^2 \omega 
   a_t^{3'}+c\, h\, W\, b_x^{2'}+
   i c	, h	,
   \kappa  b_x^{3'}+i c r^2 \omega 
   b_t^{3'}=0,
\end{multline}
and the other three are obtained by the previous exchanging $a_{\mu}^a \leftrightarrow b_{\mu}^a$, 
$W \leftrightarrow V$ and $H \leftrightarrow \Phi$.
The six linearized equation for the transverse fluctuations in the unphysical basis are:
\begin{equation}
 \begin{split}
a_y^{1''}&-c b_y^{1''}+\frac{h'
   \left(a_y^{1'}-c
   b_y^{1'}\right)}{h}-\frac{\kappa ^2
   \left(a_y^1-c b_y^1\right)}{h
   r^2}+\\
&+\frac{i \omega  \left[c (H +\Phi ) b_y^2-2 \Phi 
   a_y^2\right]+\left(\omega ^2+\Phi ^2\right) a_y^1-c
   b_y^1 \left(H  \Phi +\omega ^2\right)}{h^2}=0,
   \end{split}
\end{equation}
\begin{equation}
\begin{split}
 a_y^{2''}&-c\, b_y^{2''}+\frac{h'
   \left(a_y^{2'}-c\,
   b_y^{2'}\right)}{h}+\\
&+\frac{-i \omega  \left[c (H +\Phi ) b_y^1-2 \Phi 
   a_y^1\right]+\left(\omega^2+\Phi^2\right) a_y{}^2-c
   b_y^2 \left(H  \Phi +\omega ^2\right)}{h^2}+\\
  & +\frac{W^2
   \left(-a_y^2\right)-2 i \kappa  W a_y^3-\kappa ^2
   a_y^2+c V W b_y^2+i c \kappa  v b_y^3+i c \kappa  w
   b_y^3+c \kappa ^2 b_y^2}{h\,
   r^2}=0,
   \end{split}
\end{equation}
\begin{equation}
\begin{split}
a_y^{3''}&-c\, b_y^{3''}+\frac{h'
   \left(a_y^{3'}-c\,
   b_y^{3'}\right)}{h}+\frac{\omega ^2 \left(a_y^3-c b_y^3\right)}{h^2}+\\
&+\frac{W^2
   \left(-a_y^3\right)+2 i \kappa  W a_y^2-\kappa ^2
   a_y^3+c V W b_y^3-i c \kappa  V b_y^2-i c \kappa  W
   b_y^2+c \kappa ^2 b_y^3}{h\,
   r^2}=0,
   \end{split}
\end{equation}
and the other three are obtained by the previous exchanging 
$a_{\mu}^a \leftrightarrow b_{\mu}^a$, $W \leftrightarrow V$ and $H \leftrightarrow \Phi$.
We remind the reader that in the previous relations we have $h=h(r)=r^2-\frac{1}{r}.$

\section{Building the gauge invariant combinations of fields}
\label{gaugeinv}
In order to obtain the gauge invariant combination of the physical fields $\bar{a}_{\mu}^a$ 
and $\bar{b}^a_{\mu}$ in the condensed phase it is useful to write the transformations \eqref{1} and \eqref{2} in an explicit fashion,
\begin{equation}
\delta \begin{pmatrix}a_x^1\\ a_x^2\\ a_x^3\\ a_t^1 \\ a_t^2 \\ a_t^3 \end{pmatrix}= \begin{pmatrix} i \kappa & 0 & 0 \\ 0 & i \kappa & -W(r) \\ 0 & W(r) & i \kappa \\ - i \omega & - \Phi(r) & 0 \\ \Phi(r) & - i \omega & 0 \\ 0 & 0 & -i \omega \end{pmatrix} \times \begin{pmatrix}
\alpha^1 \\ \alpha^2 \\ \alpha^3 \end{pmatrix},
\end{equation}
\begin{equation}
\delta \begin{pmatrix}b_x^1\\ b_x^2\\ b_x^3\\ b_t^1 \\ b_t^2 \\ b_t^3 \end{pmatrix}= \begin{pmatrix} i \kappa & 0 & 0 \\ 0 & i \kappa & -V(r) \\ 0 & V(r) & i \kappa \\ - i \omega & - H(r) & 0 \\ H(r) & - i \omega & 0 \\ 0 & 0 & -i \omega \end{pmatrix} \times \begin{pmatrix}
\alpha^1 \\ \alpha^2 \\ \alpha^3 \end{pmatrix}\ ,
\end{equation}
where $\Phi(r), \; W(r), \; H(r)$ and $V(r)$ are the background fields \eqref{ans_unp_A} \eqref{ans_unp_B}.
Expressing the previous explicit relations in terms of the physical fields
\begin{eqnarray}
\label{physf1}
&\bar{a}^a_{\mu}=\frac{1}{\sqrt{2}}\sqrt{1+c}\ (a_{\mu}^a-b_{\mu}^a)\ ,\\
\label{physf2}
&\bar{b}^a_{\mu}=\frac{1}{\sqrt{2}}\sqrt{1-c}\ (a_{\mu}^a+b_{\mu}^a)\ ,
\end{eqnarray}
we obtain
\begin{equation}
\label{transform11}
\delta \begin{pmatrix}\bar{a}_x^1\\ \bar{a}_x^2\\ \bar{a}_x^3\\ \bar{a}_t^1 \\ \bar{a}_t^2 \\ \bar{a}_t^3 \end{pmatrix}= \begin{pmatrix} 0 & 0 & 0 \\ 0 & 0 & -w(r) \\ 0 & w(r) & 0 \\ 0 & - \phi(r) & 0 \\ \phi(r) & 0 & 0 \\ 0 & 0 & 0 \end{pmatrix} \times \begin{pmatrix}
\alpha^1 \\ \alpha^2 \\ \alpha^3 \end{pmatrix}\ ,
\end{equation}
\begin{equation}
\label{transform22}
\delta \begin{pmatrix}\bar{b}_x^1\\ \bar{b}_x^2\\ \bar{b}_x^3\\ \bar{b}_t^1 \\ \bar{b}_t^2 \\ \bar{b}_t^3 \end{pmatrix}= \begin{pmatrix} i \tilde{\kappa} & 0 & 0 \\ 0 & i \tilde{\kappa} & -v(r) \\ 0 & v(r) & i \tilde{\kappa} \\ - i \tilde{\omega} & - \eta(r) & 0 \\ \eta(r) & - i \tilde{\omega} & 0 \\ 0 & 0 & -i \tilde{\omega} \end{pmatrix} \times \begin{pmatrix}
\alpha^1 \\ \alpha^2 \\ \alpha^3 \end{pmatrix}\ ,
\end{equation}
where $\tilde{\omega}=\sqrt{2(1-c)}\; \omega$, $\tilde{\kappa}=\sqrt{2(1-c)}\; \kappa$,
and $\phi(r),\; w(r),\; \eta(r)$ and $v(r)$ are the physical background fields \eqref{ans_phy_A} \eqref{ans_phy_B}. 

The analysis of the gauge invariant combination of the fields $\bar{b}$ is entirely analogous to that done in Appendix A
of \cite{Gao:2012yw}, and consequently the gauge invariant combination for the fields $\bar{b}_x^3$ are:
\begin{equation}
\begin{split}
&\bar{b}_x^3+\frac{\kappa}{\omega}b_t^3+\frac{w(z) \phi(z)\bar{b}_t^1-i \omega w(z) \bar{b}_t^2}{\phi(z)^2-\omega^2}, \qquad \bar{b}_x^3+\frac{i w(z)}{\kappa}\bar{b}_x^2-\frac{w(z)^2-\kappa^2}{\kappa \omega}\bar{b}_t^3,\\
&\bar{b}_x^3+\frac{i \kappa}{w(z)}\bar{b}_x^2+\frac{i(\kappa^2  \omega-\omega w(z)^2)\bar{b}_t^2-(w(z)^2-\kappa^2) \phi(z) \bar{b}_t^1}{w(z)(\omega^2-\phi(z)^2)}\\
&\bar{b}_x^3+\frac{w(z) \phi(z)}{\kappa \omega}\bar{b}_x^1+\frac{\kappa}{\omega}\bar{b}_t^3-\frac{iw(z)}{\phi(z)}\bar{b}_t^2, \qquad \; \bar{b}_x^3+\frac{\omega w(z)}{\kappa \phi(z)}\bar{b}_x^1+\frac{\kappa}{\omega}\bar{b}_t^3+\frac{w(z)}{\phi(z)}\bar{b}_t^1 \\
&\bar{b}_x^3-\frac{(\kappa^2-w(z)^2) \phi(z)\bar{b}_x^1-i \kappa^2 \omega \bar{b}_x^2-i\kappa(\kappa^2-w(z)^2)\bar{b}_t^2}{\kappa \omega w(z)},\\
&\bar{b}_x^3-\frac{\omega(\kappa^2-w(z)^2)}{\kappa w(z) \phi(z)}\bar{b}_x^1+\frac{i \kappa}{w(z)}\bar{b}_x^2-\frac{\kappa^2-w(z)^2}{w(z) \phi(z)}\bar{b}_t^1.
\end{split}
\end{equation}
Notice that only the first of the previous combination is in general well defined in the limit of null momentum and vanishing condensates. 
As regards the gauge invariant combinations for the fields $\bar{a}$, it is easy to see from the gauge transformations 
\eqref{transform11} that there are three gauge invariant combinations of fields:
\begin{equation}
\begin{split}
&\hat{\bar{a}}_x^1=\bar{a}_x^1,\\
&\hat{\bar{a}}_t^3=\bar{a}_t^3,\\
&\hat{\bar{a}}_x^3=\bar{a}_x^3+\frac{w}{\phi}\bar{a}_t^1.
\end{split}
\end{equation}
The derivation of the gauge invariant combination of fields in the normal phase is identical to the one just outlined for the condensed phase.

\section{Phase diagrams for different values of \texorpdfstring{$c$}{}}
\label{phadif}
\begin{figure}[ht]
\centering
\includegraphics[width=90mm]{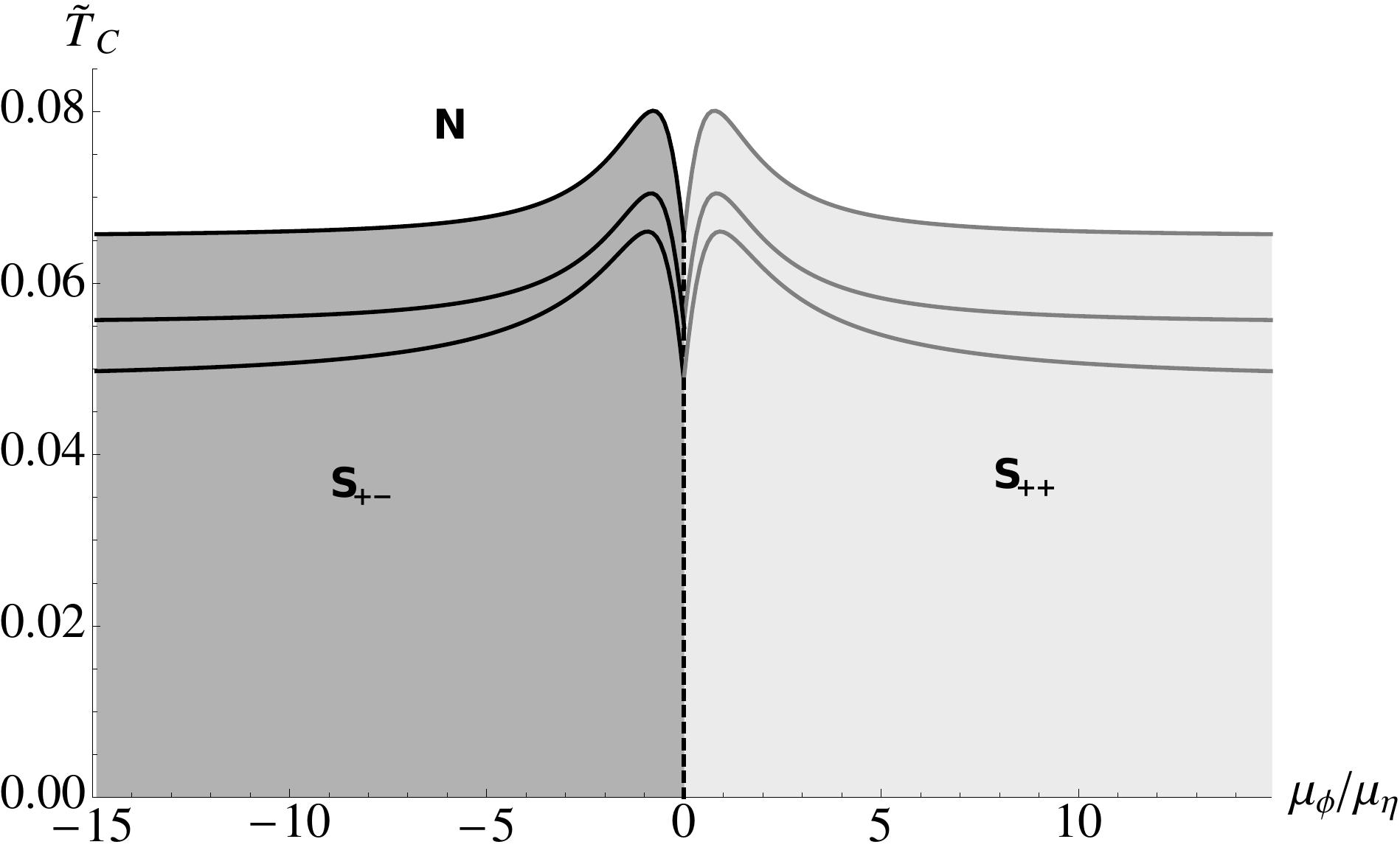} 
\caption{Phase diagrams different values of the coupling constant $c$.
We have respectively $c=1/10$ (lower) to $c=3/10$ (middle) and to
$c=5/10$ (upper). Notice that the feature of the phase diagram are qualitatively 
analogous for different values of $c$. The position of the maximal $\tilde{T}_c$ is 
weakly $c$ dependent in the range considered here.}
\label{phase_comp}
\end{figure}

\end{document}